\documentclass[aps,prd,onecolumn,nofootinbib,notitlepage,supersciptaddress,floatfix]{revtex4-1}
\usepackage[colorlinks=false, linkcolor=black, citecolor=black]{hyperref}
\usepackage{graphicx}
\usepackage[utf8]{inputenc}
\usepackage{subfig}
\usepackage{amstext,amsmath,amssymb,amsfonts,amsthm}
\usepackage[format=plain,justification=centerlast]{caption}

\def\app#1{{Appendix~\ref{#1}}}
\def\sec#1{{Section~\ref{#1}}}
\def\eq#1{{Eq.~(\ref{#1})}}

\def\fig#1{{Fig.~\ref{#1}}}

\def\dd{\mathrm{d}}
\def\be{\begin{equation}}
\def\ee{\end{equation}}
\def\ber{\begin{eqnarray}}
\def\eer{\end{eqnarray}}
\def\ba{\begin{align}}
\def\ena{\end{align}}
\def\bwt{\begin{widetext}}
\def\ewt{\end{widetext}}

\def\bra{\langle}
\def\ket{\rangle}
\def\f{\frac}

\def\pp{\partial}
\def\nn{\nonumber}

\def\tbf#1{{\textbf{#1}}}
\def\inf{\mathrm{inf}}

\def\calS{{\mathcal S}}

\begin{document}

\title{Cosmological Particle Creation Beyond de Sitter}
\date{\today}
\author{Sujoy K. Modak}
\email{smodak@ucol.mx}
\affiliation{Facultad de Ciencias - CUICBAS, Universidad de Colima, Colima, C.P. 28045, M\'exico}

%\date{}

%%%%%%%%%%%%%%%%%%%%%%%%%%%%%%%%%%%%%%%%%%%%%%%%%%%%%%%%
\begin{abstract}
Over the years, de Sitter spacetime has been a central focus, in studies involving quantum fields, for its importance in the early and late expansion stages of the universe. While de Sitter spacetime closely mimics characteristics of the inflationary and dark energy dominated universe it does not help to understand the radiation and matter dominated expansions. In this review, we revisit some recent works, involving the author, which study gravitational particle creation beyond the de Sitter stage.  Specifically, we present novel aspects of particle creation in the radiation dominated universe, and then provide an analysis of time evolution of the primordial (Bunch-Davies) vacuum state, its particle excitation and quantum versus classical behavior of field modes, in a multi stage universe, comprising of, (i) the inflationary de Sitter, (ii) the radiation dominated  and, (iii) the late dark energy dominated stages. 
\end{abstract}

\maketitle

%%%%%%%%%%%%%%%%%%%%%%%%%%%%%%%%%%%%%%%%%%%%%%%%%%%%%
%\maketitle
\tableofcontents

\section{Introduction}
In last five decades or so physicists have made considerable progress in the understanding of field theory in curved spacetime, both within the classical and quantum context. While in the classical context things are more straightforward it is not so simple for the quantum counterpart since it brings new concepts and subtleties. For example, even at the very basic level, the definition of {\em unique vacuum state} aka quantum field theory in flat spacetime looses its meaning, and a new concept of observer dependent {\em non-unique} vacuum state appears in a curved space setting. Vacuum state in standard quantum field theory in flat space is invariant under the Poincar\'e group operation. Poincare group being the group of Minkowski spacetime isometries, the inertial observers do agree with this unique definition of vacuum since they are related by Lorentz transformation. On the other hand Lorentz invariance is replaced by general covariance (diffeomorphism invariance) in a curved space and naturally Lorentz invariance of vacuum state has no such elaborate meaning since there is no natural notion of inertial observer in a curved space. Vacuum state with respect to a particular frame of reference will not necessarily be the vacuum state with respect to another frame, especially because, these two frames are generally non-inertial if we are in a curved space. In addition to this, dynamical spacetimes, where  background metrics are changing with time, add another flavor to this setting by inducing an effect of changing spacetime on the quantum state of the field itself. That is, a vacuum state at an instant may get excited in the future due to the dynamics of the spacetime itself. In this case one may also bring an instantaneous definition of vacuum state valid for a given instant of time and study how this state gets excited to show a behavior of a particle state. Therefore, in a curved space, even before considering interacting fields we get a rich characteristic of vacuum excitations just by non-trivial geometrical settings and in some cases by the dynamics of background geometry.

A great deal of interest in the study of quantum field theory in curved spacetime was generated after the seminal work by Hawking in the context of black hole evaporation \cite{hawk1, hawk2}. In fact, it was not Hawking but Parker who first gave a serious thought on particle creation in a gravitational (cosmological) framework \cite{Parker:1968mv} - \cite{Parker:1968mv2}. After these initial works, the following decades were exciting years where much of the advancements were made in all fronts. Some foundational works include (but of course not limited to) the discovery of particle creation observed by accelerating observers in flat space (the Unruh effect) \cite{unruh},  in de Sitter \cite{Mott, GH, ford, dsrev1, dsrev2, Singh:2013pxf} and anti-de Sitter spaces \cite{ads}, de Sitter in transition to radiation dominated era \cite{Hashiba:2018iff} and in more general contexts \cite{others2} - \cite{others10}, as well as, various approaches to deal with this phenomena such as the effective action formalism \cite{hu, buch}, algebraic approaches \cite{ash, Kay:1988mu, Hollands:2004yh, hack}, the tunnelling \cite{Parikh:1999mf} - \cite{Parikh:1999mf3} and complex path methods \cite{paddy} - \cite{paddy4}, and so on. In fact, in recent times,  studies on particle creation have generated an intense debate on the stability of the de Sitter space (even efforts have been made considering it as one guiding principle for resolving the cosmological constant problem). Broadly speaking, there have been proposals that the de Sitter vacuum would decay making the spacetime unstable \cite{usds} - \cite{usds13}, while a contrary point of view suggesting the opposite has also been put forward \cite{sds} - \cite{sds5}. A natural quest of understanding the cosmological constant, and hence together the inflationary universe and dark energy problems, has made the studies sharply aligned towards the de Sitter case, and the aim of this review is to go beyond this and show a rich structure of cosmological particle creation in other cosmological epochs.

There exist various elaborate studies on quantum field theory in curved space which are adequately reviewed in several articles \cite{DeWitt:1975ys, Brout:1995rd, Brandenberger:1984cz, rev-dun, rev-pad1, rev-pag1, rev-muk, rev-mar} and monographs \cite{birrel, Fulling:1989nb, book-pad, Mukhanov:2007zz, parker, paddy2}. In this review article we do not repeat the topics which have already been discussed in those works by several distinguished authors, rather, our aim here is to limit ourselves to the selected topics of particle creation, in the cosmological context, only involving recent collaborative work of the author.

As our playground, we shall consider our universe, post quantum gravity region, as described by the Friedman-Robertson-Walker (FRW) spacetime with multiple expansion stages such as - (i) inflationary stage, (ii) radiation dominated stage, (iii) matter dominated stage and (iv) dark energy dominated stage. The spacetime metric, considering a spatially flat universe (with $k=0$),  has a general form
\be
ds^2 = dt^2 - a^2(t) d\bf{x}^2
\label{frw}
\ee
where the scale factor $a(t)$ is a function of the (comoving) time $t$ and the exact form of this factor depends  on the particular expansion phase.

This article is organized as follows - in section II we study a curious example of particle creation in the radiation dominated stage. We shall install a new set of coordinates that express this spacetime as the spherically symmetric, inhomogeneous spacetime. Apart from the discussion of various observer dependent horizons we shall discuss a particle creation process which is very similar to the Unruh effect. In section III we shall move into another issue of discussing particle creation in a three stage universe (modulo the matter dominated stage). A detailed account of how the initial Bunch-Davies vacuum state is excited due to the dynamics of the spacetmime will be presented. Further connection of the results from this toy model with our understanding on the generation of initial density perturbations will be put forward.

%%%%%%%%%

\section{Particle creation in radiation dominated universe}
As found by Unruh \cite{unruh}, for an accelerated observer in Minkowski spacetime, the Minkowski vacuum state appears to be a thermal state, depends on the fact that the vacuum state corresponding to the accelerated frame (Rindler vacuum) is unitarily inequivalent to this vacuum, which is a necessary requirement for particle creation due to observers' own motion. This discovery is often regarded as one of  the most important landmarks in quantum field theory in curved space. Two most important ingredients behind Unruh effect can be identified as - (i) geometric, that is, the existence of Minkowski and Rindler spacetimes with certain symmetries which then allows the second ingredient to exist, which is, (ii) field theoretic, that is unitarily inequivalent field quantizations of scalar fields and the existence of two well defined vacuum states.

Below we present a very similar observation, albeit with important differences,  with the radiation dominated universe which was reported earlier in \cite{modak1, modak2}.

\subsection{Spherically symmetric radiation dominated universe }
This part discusses the adequate geometric  setting which is necessary for the particle creation process. We start by expressing the maximally symmetric (spatially flat) FRW metric \eqref{frw} in a spherical polar coordinate system, given by
\begin{eqnarray}
ds^2= dt^2 - a^2(t)[dr^2 + r^2(d\theta^2 +{\sin^2{\theta}}~ d\phi^2)].
\label{frw2}
\end{eqnarray}
The only unknown in the above equation is the scale factor $a(t)$ which takes different forms depending upon the epoch we are interested in. In the inflationary and dark energy dominated epochs $a(t)$ is an exponential function of time with two different exponents (inflation and dark energy driven Hubble constants) and these cases can be explained by de Sitter spacetimes (which requires a positive cosmological constant). Further, for other two epochs we have $a(t) \propto t^{n}$, that is, the scale factor has a power law behavior for radiation (for which $n=1/2$) and matter ($n=2/3$) dominated epochs. Of course these expressions for the scale factor can be derived from Einstein-Friedmann equations with appropriate entries in the energy-momentum tensor for various epochs.

Let us now consider the conformally flat form of \eqref{frw2} using the cosmological time $\eta = \int \frac{dt}{a(t)}$
\begin{eqnarray}
ds^2=a^2(\eta)[d\eta^2 - dr^2 - r^2(d\theta^2 +\sin^2{\theta}~ d\phi^2)].
\label{cffrw}
\end{eqnarray}
In the light-cone gauge, $u = \eta-r$, $v=\eta+r$ ($r=\frac{v-u}{2}$), it takes the form
\begin{equation}
ds^2 = a^2 du dv - \frac{(v-u)^2}{4}a^2 (d\theta^2 + \sin^2\theta d\phi^2).\label{ncf}
\end{equation}

\noindent It was shown in \cite{modak1} that, if we make a power law transformation of above null coordinates with arbitrary real exponent, it  does not lead us to a new spacetime with important symmetries, in general, but there is an exception for the radiation dominated universe where the new spacetime becomes (a) conformally static and (b) spherically symmetric. This aspect is discussed in appendix \ref{newst} which basically shows that a conformal transformation of the null coordinates $u$ and $v$, of the form 
\ber
U=\pm \frac{{\cal H}e}{2}u^2 \label{Uu-n}\\
V=\frac{{\cal H}e}{2}v^2 \label{Vv-n}
\eer
where $+(-)$ sign stands for $u>0 (u<0)$, take us to the new metric
\begin{equation}
d{s}^2 =  F(U,V) dU dV -    G(U,V) d\Omega^2 \label{newrd}
\end{equation}
with
\begin{eqnarray}
F(U,V) = \frac{(\sqrt{V} \pm \sqrt{\pm U})^2}{4\sqrt{\pm UV}},\\
G(U,V) = \left(\frac{V- (\pm U)}{2}\right)^2.
\end{eqnarray}
where `$+$' and `$-$' signs now are applicable for $U>0$ or $U<0$, respectively. 

Now, using new time and radial coordinates
\begin{eqnarray}
T = (V+U)/2\label{RTN2}\hspace{1cm}; \hspace{1cm} R= (V-U)/2. \label{RTN1}
\end{eqnarray}
the spacetime \eqref{newrd} for $U \ge 0$ or $T\ge R$, which we call  Region-I, can be expressed as
\be
d{s}_I^2 = F_I(T,R) (dT^2 -dR^2) - R^2 d\Omega^2 \label{mRT00}
\ee
with
\be
F_I(T,R) = \frac{(\sqrt{T+R} + \sqrt{T-R})^2}{4\sqrt{T^2 - R^2}}. \label{FTR1}
\ee
Whereas, for Region II  ($U \le 0$ or $T\le R$), we get
\be
d{s}_{II}^2 = F_{II}(T,R) (dT^2 -dR^2) - T^2 d\Omega^2 \label{mRT11}
\ee
with
\be
F_{II}(T,R) = \frac{(\sqrt{R+T} - \sqrt{R-T})^2}{4\sqrt{R^2 - T^2}}. \label{FTR2}
\ee
 
\noindent Notice that, in region I (\eqref{mRT00}) we have 
\ber
T &=& (V+U)/2 = \frac{{\cal H}e}{2}(\eta^2 + r^2) \label{RT2}\\
R &=& (V-U)/2 = {\cal H}e \eta r \label{RT1},
\eer
and for region II, the relationships between the two sets of coordinates are reversed, so that
\ber
T &=& (V+U)/2 ={{\cal H}e \eta r} \label{ter}  \\
R &=& (V-U)/2 =  \frac{{\cal H}e}{2}(\eta^2 + r^2) \label{uf1}\label{rer}.
\eer

\noindent We can also express the conformal factor $F(T,R)\rightarrow F(H(T,R),R)$ (where $H(T,R)$ is the Hubble parameter in the radiation epoch) and express the intervals \eqref{mRT00} and \eqref{mRT11} 
 %\bwt
 \ber
ds_I^2 &=& \frac{dT^2 - dR^2}{1-H^2R^2} -R^2 (d\theta^2 + \sin^2\theta d\phi^2) ,~~\text{for}~~R\le1/H \label{mRT2}
\eer
%\ewt
which defines the region I; and 
%\bwt
\ber
ds_{II}^2       &=& \frac{dT^2 - dR^2}{H^2T^2 -1} -T^2 (d\theta^2 + \sin^2\theta d\phi^2),~~\text{for}~~R\ge1/H, \label{mRT3}
\eer
%\ewt
which now defines the region II. It is clear from these two expressions that region I covers the sub-Hubble region while region II covers the super-Hubble region. Therefore, the union of two covers the full spacetime describing the radiation dominated universe. For the future convenience, and for avoiding a confusion among various coordinates appearing in this review, we shall refer the above \eqref{mRT2} and \eqref{mRT3} as $(T,R)$ metric and the coordinate system as $(T,R)$ coordinates. We are naming this just because of clarity in presentation which will be needed when we switch between various frames - such as Minkowski, Rindler, cosmological and the above.

It is important to note that $T$ is always timelike and $R$ is always spacelike since there is no signature change between \eqref{mRT2} and \eqref{mRT3}, while passing from sub to super Hubble scale. The entire spacetime which was originally defined by the cosmological coordinates with coordinates $ 0< \eta<\infty$ and $0 <r <\infty$ in \eqref{cffrw} is now covered by new coordinates $0 <T < \infty$ and $0<R<\infty$, and jointly by \eqref{mRT2} and \eqref{mRT3}  (or by \eqref{mRT00} and \eqref{mRT11}). In original cosmological coordinates the radiation stage Hubble parameter $H$ is only a function of cosmological time $\eta$ (or comoving time $t$), however, in new coordinates $H$ is a function of both $T$ and $R$. Therefore the size of the observable  universe is different for different observers, situated at various spatial points. The new metrics,  \eqref{mRT2}, \eqref{mRT3},  are isotropic (because of the spherical symmetry) but  they are inhomogeneous since the metric coefficients not only depend on $T$ but also on $R$. Although an observer located at small $R<<1/H$ may ignore the quadratic term as compared to 1 in \eqref{mRT2} and the metric becomes homogeneous and isotropic for her (i.e.,  maximally symmetric just like the FRW metric). However, as she approaches the  Hubble scale $R\sim1/H$ metric becomes highly inhomogeneous. This is a direct consequence of the conformal transformation which does not respect the homogeneity. The radius of two sphere in \eqref{mRT2} is spacelike, implying that the size of the universe does not change at all for a static observer at a constant $R$; however, in \eqref{mRT3} the radius is timelike and in fact given by $T$, implying the radius itself is a measure of time.

\subsection{Observers and horizons}
Here we shall have a closer look on various observers using $(T,R)$ metrics. In fact, we shall distinguish observers in two categories - (i) static and (ii) nonstatic in $(T,R)$ frame. 

\begin{figure}[t]
\centering
%\rotatebox{270}{
\includegraphics[angle=0,width=10cm,keepaspectratio]{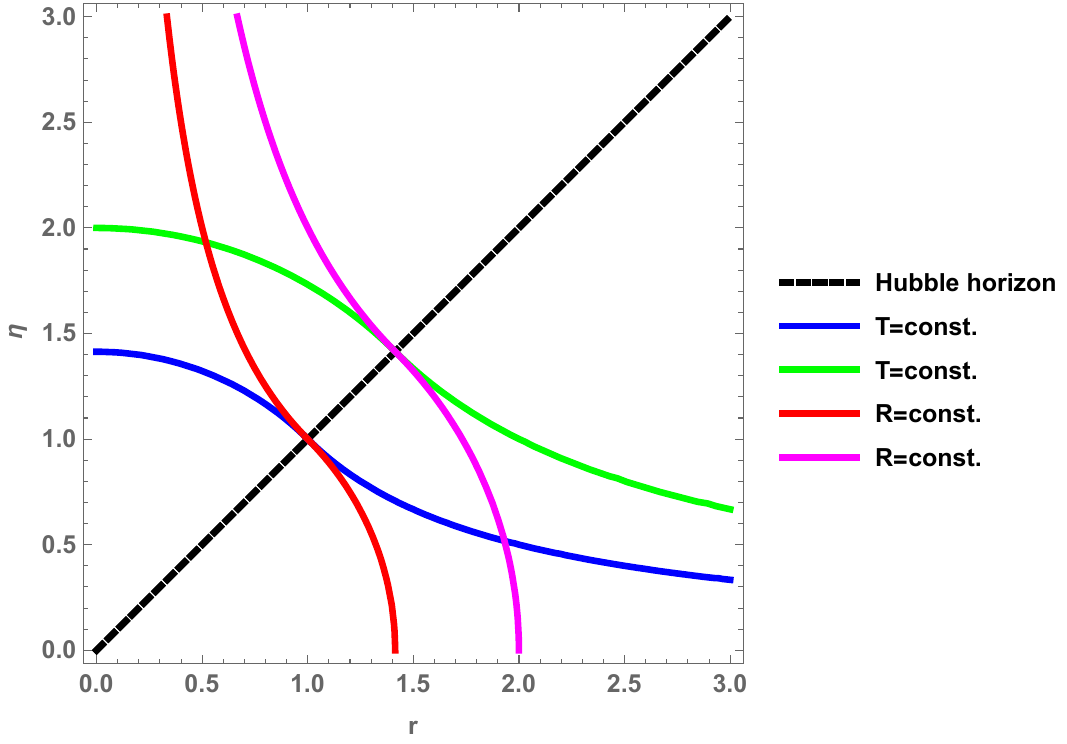}
\caption[]{$T$ and $R$ constant curves in ($\eta,~r$) plane. The intersection points are on the light-cone boundary at Hubble scale. Note that $T$ always remain timelike, $R$ always remain spacelike. Any static observer at $R=const.$ can get causal information from any point in the full spacetime. There is no horizon for a static observer at the Hubble scale. However, non-static observers find a horizon.}
\label{fig-1}
\end{figure}

Let us first consider the static observers at a constant $R$. From \eqref{RT2} and \eqref{RT1} we find that $T=const.$ and $R = const.$ curves in the ($\eta,~r$) plane are given by the circular and linearly inverse trajectories in region-I.  These are just opposite  in region-II. They are plotted in Fig. \ref{fig-1} and we note that any $R=const.$ trajectory in asymptotic past ($\eta \rightarrow 0$) and in asymptotic future ($\eta \rightarrow \infty$) are essentially indistinguishable from the freely falling observers ($r=$ const) in cosmological frame - the inertial observer in cosmological frame, defined at infinite past, starts accelerating and attains luminal velocity, momentarily, near the Hubble radius, and then decelerates which make the observer to cross the Hubble radius, from super-Hubble to sub-Hubble region. Then this observer keeps decelerating (in an identical rate of acceleration at super Hubble scale) and finally becomes freely falling, once again, moving towards the centre $r=0$, at asymptotic future. In summary, the initial freely falling observer finally becomes freely falling - the difference being they may start from any value of $r$ but they all end up towards $r=0$ asymptotically. It is important to mention that {\em the static observers can obtain signal from anywhere in the spacetime - there appears no horizon as the worldline remain geodesically complete}. This is also reflected in no signature change in the metric \eqref{mRT00} and \eqref{mRT11}. The static observer while approaching the horizon, as evident from Fig. \ref{fig-1}, follows the trajectory $\eta + r = 0$ (or $T + R=0$). Along this line $F_I$ and $F_{II}$ diverges, however, since we also have $dT=-dR$, the factor multiplying $F_I$ and $F_{II}$ vanishes, and this makes the interval $ds^2$ finite. 

Now let us consider the non-static observers in $(T,R)$ frame whose trajectory can be expressed as $T = G(R)$. Along this worldline the ($R,T$) sector of the metric \eqref{mRT00} becomes $ds^2 = \Xi  dR^2$, where the conformal factor
\be
\label{xi}
\Xi = \frac{(\sqrt{G(R)+R} + \sqrt{G(R)-R})^2}{4\sqrt{G(R)^2 - R^2}} \times (G'^2(R) -1).
\ee
If this factor diverges, for some allowed value of $R>0$ and for a given $T=G(R)$, it will indicate the presence of horizon for the observer satisfying the aforementioned trajectory. We are now free to select any observer trajectory and verify if the horizon appears or not. 

Take for example a linear trajectory $G(R) = \alpha_0 R + \beta_0$, where $\alpha_0 \ge 1$ (necessary for region I) and $\beta_0 >0$ are constants with appropriate dimensions. This will mean that the radial velocity of the observer is constant $dR/dT ={1}/{\alpha_0}$ and therefore they have no radial acceleration. In this case, it is easy to check that $\Xi$ does not diverge. {\it Therefore, observers with constant radial velocity in $(T,R)$ frame do not encounter a horizon}. This is slightly analogous to the result in Minkowski spacetime where inertial observers with constant four velocity do not encounter a horizon. In our case, these observers can be regarded as the freely falling observers, without experiencing a force field, in $(T,R)$ frame. Same conclusion can be reached by inspecting \eqref{mRT11}.

Now, let us consider a quadratic equation, $G(R) = \alpha_1 R^2 + \beta_1$, where $\alpha_1$ and $\beta_1$ are dimension-full positive definite constants. These observers have a constant radial acceleration ${1}/{\alpha_1}$ and follow a parabolic trajectory in $(T,R)$ plane. Substituting this in \eqref{xi} we find
\ber
\label{xir}
&&\Xi \notag \\
&&= \frac{(2\alpha_1 R+1)(\sqrt{(\alpha_1 R +1) R + \beta_1} + \sqrt{(\alpha_1 R +1) R - \beta_1})^2}{4\sqrt{\alpha_1 R^2 + \beta_1 + R}}   \nonumber \\
&&\times \frac{2\alpha_1 R -1}{\sqrt{\alpha_1 R^2 + \beta_1 - R}}.
\eer
The first factor is always finite, so it is only the second factor which determines if $\Xi$ diverges. Clearly, there are two divergences for the root $R_0 = \frac{1\pm\sqrt{1-4\alpha_1\beta_1}}{2\alpha_1}$, if $\beta_1<1/4\alpha_1$. So there are two horizons, in general, for this observer with constant radial acceleration.  However, if $\beta_1$ vanishes then there is only one horizon at $R_0 = 1/\alpha_1$. Further, there is a very special case for  $\beta_1 = 1/4\alpha_1$ for which the second factor in \eqref{xir} is $\sqrt{4\alpha_1}$. Therefore, {\it for this and only this case, observers with a constant radial acceleration do not encounter a horizon}. What is more interesting, is to show that {\it this observer is none other than the so called Fundamental Cosmological Observer (FCO)}. FCOs are the {\it static observers} in cosmological frame situated at constant $r$. As we see here,  they are special class of observers in $(T,R)$ frame who are accelerating radially and yet have a full access to the spacetime - this is of course expected because FCOs are the static observers in the original FRW metric, in cosmological frame, who do not encounter a horizon either.  

To demonstrate this we use relations \eqref{RT2} and \eqref{RT1}. We find that the constant $\eta=\eta_0$ trajectories in $RT$ plane satisfy an identical relationship like $T= \alpha_1 R^2 + \beta_1$ with $\beta_1 = 1/4 \alpha_1 =  {\cal H} e \eta_0^2/2$. In fact a constant $r$ trajectory also satisfies the same relationship just because \eqref{RT2} and \eqref{RT1} are symmetric under the interchange of $r$ and $\eta$. Therefore, FCOs do not encounter a coordinate singularity. It therefore needs no more justification to say that neither the comoving observers with proper time $t$ will encounter a horizon in $(T,R)$ spacetime. One can, in fact, go on to discuss other observer trajectories but we rather want concentrate on the FCOs as our case study due to their physical importance in understanding the cosmos.

\begin{figure}[t]
\centering
%\rotatebox{270}{
\includegraphics[angle=0,width=8cm,keepaspectratio]{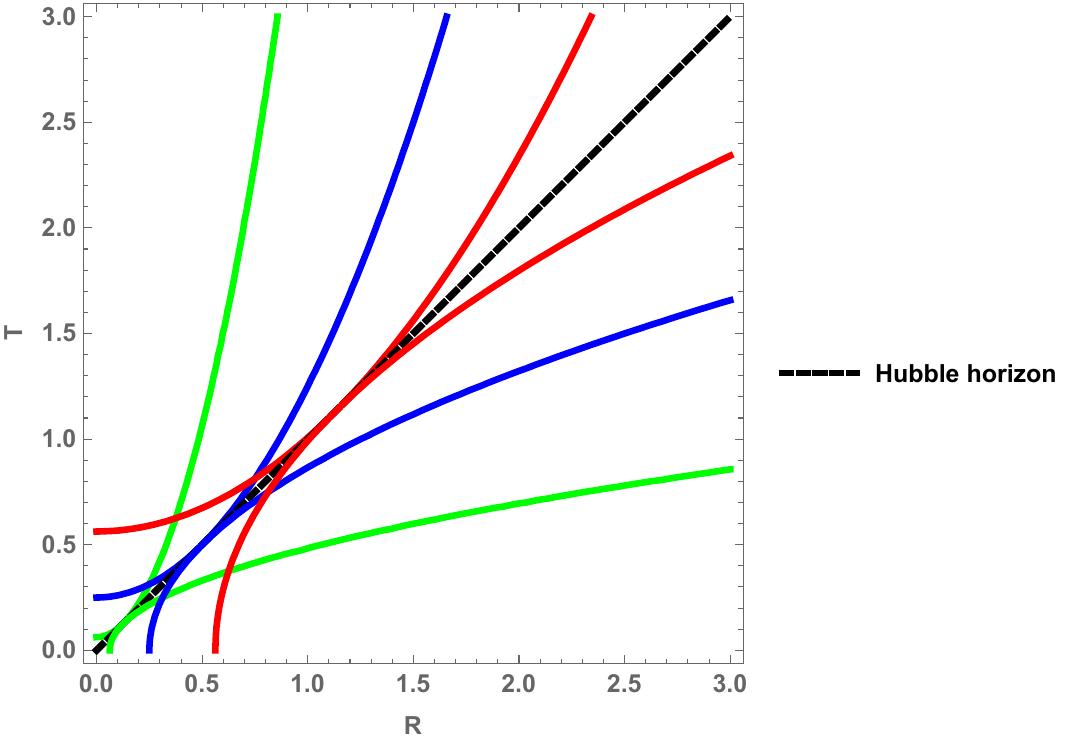}
\caption[]{Constant time and space slices in cosmological frame depicted in the $(T,R)$ plane. The black dotted line is the comoving Hubble radius. Different pairs of $r=$ const. and $\eta=$ const. slices are in same color. For example, the green horizontal curve is a spacelike hypersurface (or a ``time-slice'') corresponding to a constant $\eta$ and the vertical one is a timelike hypersurface (or a ``space-slice'') corresponding to a constant $r$. Same convention applies to others. The intersection points are on the light-cone boundary at the Hubble scale. There is no horizon for FCOs at the Hubble scale.}
\label{fig1}
\end{figure}

Particularly, we want to see how the constant $\eta$ (which defines a ``time-slice'') and constant $r$ (which defines a ``space-slice'') orbits will look like. Since the relationships between the two sets of coordinates, in \eqref{RT2} and \eqref{RT1}, are symmetric under the exchange of $\eta$ and $r$ (so is true for \eqref{ter} and \eqref{rer}) we have to be rather cautious to identify the time-slices and the space-slices. The timeslices ($\eta= \eta_0$), which needs to be forward moving with $T$, as $\eta$ increases, are defined as follows
\begin{multline}
\left.
\begin{array}{ll}
 T = \frac{R^2}{2 {\cal H} e \eta_0^2} + \frac{1}{2}{\cal H} e \eta_0^2  & \hspace{10pt}R \le T \\
        R = \frac{T^2}{2 {\cal H} e \eta_0^2} + \frac{1}{2}{\cal H} e \eta_0^2 & \hspace{10pt}R \ge T \\
\end{array}
\right\}
\label{tconst}
\end{multline}
whereas, for spaceslices ($r =r_0$ ) these are reversed:
\begin{multline}
\left.
\begin{array}{ll}
 R = \frac{T^2}{2 {\cal H} e r_0^2} + \frac{1}{2}{\cal H} e r_0^2  & \hspace{10pt}R \le T \\
        T = \frac{R^2}{2 {\cal H} e r_0^2} + \frac{1}{2}{\cal H} e r_0^2 & \hspace{10pt}R \ge T \\
\end{array}
\right\}
\label{rconst}
\end{multline}
These slices foliate the complete spacetime, (basically the $(T,R)$ plane; each point in this plane is a two sphere) with constant $\eta$ (timeslices) and constant $r$ (spaceslices), in a consistent manner so that the Cauchy problem is well posed. This is shown pictorially through the plots in Fig. \ref{fig1}. In summary, Fig. \ref{fig-1} and Fig. \ref{fig1} provide two different foliation of the full spacetime by means of spacelike and timelike hypersurfaces. Unlike, Rindler and Minkowski case, here cosmological and $(T,R)$ coordinates cover the complete patch of the spacetime and hence when we make field quantization the mode functions will be complete for both cases. This does not happen in Rindler-Minkowski case since the Rindler observers only cover one fourth of the spacetime due to coordinate singularity and therefore the mode functions in Rindler frame remain incomplete. But here, neither FCOs nor static observers in $(T,R)$ frame encounter a horizon. This will have an important impact on the particle creation process. 

Notice that, once again, in Fig. \ref{fig1} the FCOs at $r=$const. (vertical curves) are in fact freely falling, in $(T,R)$ frame only in the asymptotic past and future, but they are accelerated (and decelerated) radially  in super (and sub) Hubble regions, respectively. FCOs do attain a  luminal velocity at Hubble scale and this is exactly analogous to the case of static $(T,R)$ observers in cosmological frame. In fact, we expect this pattern to be reciprocal because they are, after all,  accelerated or decelerated with respect to each other. Another important thing to note that,  an initial Cauchy data, on an initial space-like hypersurface, may be defined either on a $T=$ const surface in Fig. \ref{fig-1} or on a $\eta=$ const surface in Fig. \ref{fig1}. Both these datasets, given in any of the initial hypersurfaces, will be satisfactorily time transported to future hypersurfaces by two different time translations. 

\subsection{Particle creation: two dimensional analysis}
Let us now proceed with our main goal of discussing cosmological particle creation in the radiation dominated universe. We shall do it separately for two and four spacetime dimensions.

\subsubsection{Static observer in $(T,R)$ frame}
The two dimensional toy version is much simpler but very useful to get physical insights. The massless Klein-Gordon equation for the background \eqref{ncf} and \eqref{newrd} (by ignoring $\theta,~\phi$ part) becomes
\begin{eqnarray}
\partial_u\partial_v \Phi= 0,\\
\partial_U\partial_V \Phi=0
\end{eqnarray}
respectively. The solutions are just plane-wave modes which leads to following expansions of the field operator 
\ber 
 \hat{\Phi} &=& \int_o^{\infty} \frac{d\omega}{\sqrt{4\pi\omega}}(e^{-i\omega u} a_\omega + e^{i\omega u} a_\omega^{\dagger} + \text{right moving}) \nonumber\\ \label{exp1}\\
 &=& \int_o^{\infty} \frac{d\omega}{\sqrt{4\pi\Omega}}(e^{-i\Omega U} b_\Omega + e^{i\Omega U} b_\Omega^{\dagger} + \text{right moving}) \nonumber\\ \label{exp2}.
\eer 

Notice that there appear two sets of creation and annihilation operators which in fact define two different vacuum states. These are defined as follows which we term as the cosmological vacuum ($|0_{\text{C}}\rangle$) and $T-$vacuum ($|0_{\text{T}}\rangle$) (once again for the sake of clarity)
\ber
a_\omega |0_{\text{C}}\rangle = 0 \\
b_\Omega |0_{\text{T}}\rangle = 0.
\eer 

Bogolyubov transformation for the left moving modes (calculations for right moving modes are fully analogous) yields
\be
b_\Omega = \int_0^{\infty} d\omega (\alpha_{\Omega\omega} a_\omega - \beta_{\Omega\omega} a_\omega^{\dagger}).\label{bt}
\ee
Substituting this in \eqref{exp2} and comparing we get
\be
\frac{1}{\sqrt{\omega}} e^{-i\omega u} = \int_0^\infty\frac{d\Omega'}{\sqrt{\Omega'}} \left(\alpha_{\Omega'\omega} e^{-i\Omega' U} - \beta_{\Omega'\omega} e^{i\Omega' U}\right).
\ee
Multiplying both sides with $e^{\pm i \Omega U}$ and integrating we obtain
\ber
\alpha_{\Omega\omega} &=& \frac{1}{2\pi}\sqrt{\frac{\Omega}{\omega}} \int_{-\infty}^{\infty} dU e^{-i\omega u +i\Omega U}, \\
\beta_{\Omega\omega} &=& -\frac{1}{2\pi}\sqrt{\frac{\Omega}{\omega}} \int_{-\infty}^{\infty} dU e^{i\omega u +i\Omega U}.
\eer
Since we have $U=\pm\frac{{\cal H}e u^2}{2}$ ($`+'$ for $U\ge 0$ and $`-'$ for $U\le 0$) we can express
%\bwt
\ber
\beta_{\Omega\omega} &=& - \frac{{\cal H}e}{2\pi}\sqrt{\frac{\Omega}{\omega}} \left( \int_{0}^{\infty} (u du) e^{i\omega u + i\Omega ({\cal H}e u^2/2)} + \int_{-\infty}^0 (-u du) e^{i\omega u -i\Omega ({\cal H}e u^2/2)}\right),\\
&=&  I + I^* \label{bog2d}
\eer
%\ewt
 The first integral can be written in the following form
\be 
I =  - \frac{{\cal H}e}{2\pi}\sqrt{\frac{\Omega}{\omega}} e^{-i\frac{\omega^2}{2{\cal H}e\Omega}}\int_0^\infty u du ~e^{\frac{i{\cal H}e \Omega}{2}(u+\frac{\omega}{{\cal H}e \Omega})^2}.
\ee
Making a change of variable by defining $x=\frac{{\cal H} e \omega}{2}(u+\frac{\omega}{{\cal H}e\Omega})^2$ we get
%\bwt
\be 
I = -\frac{1}{2\pi} \sqrt{\frac{{\cal H}}{2 \omega}}  e^{({\frac{1}{2} - \frac{i \omega^2}{2{\cal H}e \Omega}})} \left(\sqrt{\frac{2}{{\cal H} e \Omega}} \int_{\frac{\omega^2}{2 {\cal H} e \Omega}}^\infty {dx~ e^{i x}}  - ({\omega}/{{\cal H}e \Omega})  \int_{\frac{ \omega^2}{2 {\cal H} e \Omega}}^\infty {dx ~x^{-1/2} ~e^{ix}} \right)
\ee
%\ewt
This kind of integral, often encountered in the calculation of Bogolyubov coefficients, is calculated by using the identity \cite{paddy2}
%\bwt
\be
\int_{x_0}^{\infty} dx x^{s-1} e^{-bx} = e^{-s\log b} \Gamma[s, x_0];~~ Re~ b, Re~ s >0.
\label{id}
\ee
%\ewt
where $\Gamma[s, x_0]$ is an upper incomplete gamma function. Using this and further using the relation $\lim_{\epsilon\rightarrow 0}\log|{\cal H}e \Omega/2i +\epsilon| \simeq \log| {\cal H}e \Omega/2| - \frac{i \pi}{2}$ (using $\Omega >0$), we obtain
%\bwt
\be
I = -\frac{1}{2\pi} \sqrt{\frac{{\cal H}}{2 \omega}}  e^{({\frac{1}{2} - \frac{i \omega^2}{2{\cal H}e \Omega}})} 
\left(\sqrt{\frac{2}{{\cal H} e \Omega}} ~e^{i \pi/2}~ \Gamma[1, \frac{\omega^2}{2 {\cal H} e \Omega}] - ({\omega}/{{\cal H}e \Omega})  ~e^{i \pi/4} ~\Gamma[\frac{1}{2}, \frac{\omega^2}{2 {\cal H} e \Omega}] \right),\label{2d1}
\ee
%\ewt
and  using \eqref{bog2d} we get
%\bwt
\be
\beta_{\Omega\omega} = -\frac{1}{\pi} \sqrt{\frac{{\cal H} e}{2 \omega}} \left(\sqrt{\frac{2}{{\cal H} e \Omega}} ~\sin(\frac{\omega^2}{2 {\cal H} e \Omega}) ~\Gamma[1, \frac{\omega^2}{2 {\cal H} e \Omega}] - \frac{\omega}{{\cal H}e \Omega}  ~\cos(\frac{\pi}{4} - \frac{\omega^2}{2 {\cal H} e \Omega}) ~\Gamma[\frac{1}{2}, \frac{\omega^2}{2 {\cal H} e \Omega}] \right).
\label{betah}
\ee
%\ewt
This is of course an exact expression and one can readily use it to calculate the exact particle number density as defined in \eqref{def}. However, it is quite difficult to find an analytical closed form expression for which one needs to use numerical path. 

Here we just restrict us upto a analytical expression which is only possible to get under some approximations. We consider following approximations - (i)  $\frac{\omega}{\Omega} << {\cal H}$, i.e., the ratio between the frequencies is much smaller than the inflationary Hubble constant (${\cal H}\sim 10^{37} s^{-1}$), and (ii) finite $\omega$. The  inequality $\frac{\omega}{\Omega} << {\cal H}$ is interpreted in the following manner - if we are counting the particle excitation number with frequency $\Omega \sim {\cal O}(1)$, then this inequality ensures that we do not sum over the field modes in the inflationary era for which $\omega \sim \omega_{inf}\sim 10^{37} s^{-1}$ and this serves as an ultra-violet cut-off. Therefore, this calculation will hold quite strongly while counting particle excitation for higher frequencies ($\Omega$) than compared to its lower values.  Under such a condition equation  \eqref{betah} simplifies to (the upper incomplete gamma function becomes complete, $\sin\theta$ becomes $\theta$ and the second term inside the big parenthesis in \eqref{betah} drops out altogether)
\be
\beta_{\Omega\omega} \sim - \frac{1}{2\pi {\cal H}e} \left(\frac{{\omega}}{\Omega}\right)^{3/2}.
\label{betap}
\ee
Now substituting $\frac{\omega}{\Omega}=\epsilon{\cal H}$ and using \eqref{def}
we get the particle number density
\ber
\langle n_{\Omega} \rangle &\sim&  \int_{0}^{\omega} d\omega~|\beta_{\Omega\omega}|^2 \label{def}\\
&\propto& \epsilon^3 {\cal H} \label{n2d}
\eer
Although the inflationary Hubble constant is very large (${\cal H} \sim 10^{37} s^{-1}$), since it is multiplied with $\epsilon^3$, the left hand side becomes very small. If we rather count all the particles during the entire lifetime of the radiation dominated universe this particle number density can still provide a finite energy density. A full numerical analysis would be interesting and this remains open so far.

\subsubsection{Fundamental Cosmological Observer}
Now we proceed into another important part of our discussion which is gravitational particle creation for FCOs.  As showed in Fig. \ref{fig1}, these observers have non-trivial trajectory in cosmological FRW coordinates. Here we want to calculate the particle content for FCOs who are at $r=$const. and following the trajectories showed in Fig. \ref{fig1}. These observers will find $T-$ vacuum (to be defined shortly) as a particle excited state. Therefore, FCOs will be exposed to a radiation due to their motion.

The inverse Bogolyubov transformation to \eqref{bt} relating the annihilation operator $$a_\omega = \int_0^\infty d\Omega (\alpha'_{\omega\Omega} b_\Omega - \beta'_{\omega\Omega} b_\omega^\dagger)$$ in terms of the sum of creation and annihilation operators of the other basis can be easily calculated as 
\ber
\alpha'_{\omega\Omega} &=& \frac{1}{2\pi}\sqrt{\frac{\omega}{\Omega}} \int_{-\infty}^{\infty} du e^{-i\Omega U +i\omega u}, \\
\beta'_{\omega\Omega} &=& -\frac{1}{2\pi}\sqrt{\frac{\omega}{\Omega}} \int_{-\infty}^{\infty} du e^{i\Omega U +i\omega u}.
\label{beta}
\eer
The average particle number density for a given frequency is then given by
\be
\langle n_{\omega} \rangle = \int_0^{\infty} d\Omega |\beta'_{\omega\Omega}|^2
\label{nom}
\ee
where, $n_{\omega} = a_\omega^\dagger a_\omega$ is the number operator defined in the cosmological basis and the expectation value $\langle 0_{\text{T}} | n_{\omega} | 0_{\text{T}}\rangle$ is calculated in the vacuum state in the new basis, as defined  by $b_\Omega |0_{\text{T}}\rangle = 0$. We refer to this vacuum state  $|0_{\text{T}}\rangle$ as $T-$ vacuum.

To calculate the coefficient \eqref{beta} we first divide the integral for $u \le 0$ and $u\ge 0$. After performing the integration \eqref{beta} we can derive
\be
|\beta'_{\omega\Omega}|^2 = \frac{\omega }{8 e {\cal H} \pi^2 \Omega^2 } \left(1+ \sin(\frac{\omega^2}{\Omega e {\cal H}})\right)\Gamma^2[\frac{1}{2}, \frac{\omega^2} {2 \Omega e {\cal H}}] \label{beta1}
\ee
where $ \Gamma$ is an upper incomplete gamma function. Equation \eqref{nom} then provides 
average particle number density. Unfortunately, it is difficult to get an exact analytical result for the particle number density \eqref{nom} using \eqref{beta1}. We therefore again use a similar approximation like the last subsection setting $\frac{\omega}{\Omega} << {\cal H}$ once again. However, unlike the last time, in this case this inequality is achievable only by putting a infra-red cut off on $\Omega$ while performing the integration in \eqref{nom}. Again, using this approximation we consider the incomplete gamma as complete and $\sin\theta$ as $\theta$, implying $|\beta'_{\omega\Omega}|^2 \sim \frac{\omega }{8 e {\cal H} \pi \Omega^2 } \left(1+ \frac{\omega^2}{\Omega e {\cal H}}\right)$ and therefore
\be
\langle n_{\omega} \rangle \sim \frac{\omega }{8 \pi e {\cal H}  } \left(\frac{1}{\Omega_0}+ \frac{\omega^2}{2e {\cal H} \Omega_0^2}\right)
\label{nom1}
\ee
where $\Omega_0$ is the infra-red cut off that we considered here. The number density again is quite small in this regime since $\frac{\omega}{{\cal H}\Omega_0}<<1$. Also, this calculation is just a crude approximation and a legitimate numerical analysis needs to be done to understand the situation in a fully satisfactory manner. We are in the process of a full numerical analysis with this and all other relevant cases at the time of writing this report.
%%%%%%%%%%%%%%%%%%%%%%
%%%%%%%%%%%%%%%%%%%%

\subsection{Particle creation: four dimensional analysis}
Now, we consider a realistic four dimensional setup. Here, we shall only limit ourselves to the static observers in $(T,R)$ frame who will envision the cosmological vacuum state as a particle excited state. The other important aspect involving the FCOs looking into $T-$ vacuum is under investigation at the time of writing this report.

\subsubsection{Quantum fields in cosmological frame}

Consider a massless scalar field of arbitrary coupling and solve the Klein-Gordon equation $\Box \Phi =0$  (since Ricci scalar $R$ is zero in radiation stage) in the radiation dominated universe, using the metric \eqref{cffrw}. Use a separation of variables considering this background geometry, 
\be 
\Phi (\eta, r, \theta, \phi) = \sum_{l} \frac{f_l (r)}{r} g(\eta) Y_{lm} (\theta, \Phi) \label{anphi}
\ee
where, the angular part $Y_{lm}$ represents the spherical harmonics, and, the $(\eta,~r)$ dependent parts are the solutions of the following equations
\ber
\eta^2 \frac{d^2 g}{d\eta^2} + 2\eta \frac{dg}{d\eta} + \omega^2 \eta^2 g  &=& 0, \label{etag}\\
\frac{d^2 f_l}{dr^2} + \left(\omega^2 - \frac{l(l+1)}{r^2} \right)f_l &=& 0. \label{rf}
\eer
Let us now rescale $\eta' = \omega\eta$ and express \label{etag} as a spherical Bessel equation with $n=0$
\be
\eta'^2 \frac{d^2 g}{d\eta'^2} + 2\eta' \frac{dg}{d\eta'} + \eta'^2 g  = 0.
\ee
Solution of this equation is given by the spherical Bessel function of the first kind $j_{0} (\eta') = \sin\eta'/\eta'$ and the second kind $n_0(\eta') = - \cos\eta'/\eta'$. Equivalently, we can construct the linear combinations $h^{(1/2)}_0 (\eta') = -n_{0} (\eta') \pm i j_{0} (\eta')  = {e^{\pm i \eta'}}/{\eta'} =  {e^{\pm i \omega\eta}}/{\omega\eta}$ which makes $\Phi$ the real scalar field. Moreover, if we consider only upto the s-wave approximation ($l=0$ in \eqref{rf}), we have $f_0 (r) = e^{\pm i \omega r}$, and the ansatz \eqref{anphi} can be expressed as 
\ber
\Phi &=& \sum_{\omega} {\cal N} \frac{f_0 (r)}{r} g_{\omega}(\eta) Y_{00} (\theta, \Phi) \\
&=& \sum_{\omega} {\cal N} \left( \frac{e^{-i\omega(\eta - r)}}{2\sqrt{\pi}~\omega \eta r} a_\omega^{L} + \frac{e^{-i\omega(\eta + r)}}{2\sqrt{\pi}~\omega \eta r} a_\omega^{R} \right) + \text{h.c} \label{clphi}
\eer
where, ${\cal N}$ is the normalization constant which can be determined by imposing a orthogonality condition of the inner product between field modes; superscripts $L$ and $R$ stand for the left and right moving field modes. Coefficients $a_\omega^{L/R}$ will be elevated to the annihilation operators  at the time of elevating $\Phi$ as the field operator. 

For determining the normalization constant ${\cal N}$, recall the inner product defined on the spacelike surface $\Sigma$
\be
(\Phi_{\omega_1}, \Phi_{\omega_2}) = -i\int_{\Sigma} \sqrt{-\gamma}d^3x  ~n^{\mu} (\Phi_{\omega_1} \nabla_\mu\Phi_{\omega_2}^* - \Phi_{\omega_2}^* \nabla_\mu\Phi_{\omega_1}).
\ee
where $\Sigma$ is a $\eta=const.$ hypersurface which also sets $n^{\mu} = 1/a (1,0,0,0)$. For the radiation dominated universe $a = a_0 t^{1/2} = (a_0^2/2) \eta$ and, using the mode functions in \eqref{clphi} we get (after using $a_0=\sqrt{2{\cal H}e}$)
\be
{\cal N} = \frac{\sqrt{\omega}}{\sqrt{4\pi}{\cal H}e}.
\ee
Plugging this in \eqref{clphi} and changing the convention from a summation to integration we get 
\be 
\Phi = \int {d^3\omega} \left( u_{\omega}  a_\omega^{L} + v_{\omega}  a_\omega^{R} \right) + \text{h.c}
\ee
where
\ber
u_\omega = \frac{e^{-i\omega(\eta - r)}}{4\pi {\cal H}e\sqrt{\omega}~ \eta r} \label{mod1} \\
v_{\omega} = \frac{e^{-i\omega(\eta + r)}}{4\pi {\cal H}e \sqrt{\omega}~ \eta r} .\label{mod2}
\eer
We now interpret $\Phi$  as the field operator which then defines the cosmological vacuum for left  and right moving modes as $a_{\omega}^{L} |0_{\text{C}}\rangle^{L} =0$ and $a_{\omega}^{R} |0_{\text{C}}\rangle^{R} =0$.
%\vskip 2cm

%%%%%%%%%%%%%%%%%%%%%%%%%%%%%%%%%%%%%

\subsubsection{Quantum fields in $(T,R)$ frame}
\label{subh}
First we consider the sub-Hubble region by using the massless scalar field equation in the background of metric (\ref{mRT2}), and look for a spherical wave type solution 
\begin{eqnarray}
\Phi_{\Omega}=\displaystyle{\sum_{l,m}}\frac{\Phi_{\Omega}^{lm}(T, R)}{R}Y_{lm}(\theta,u).
\end{eqnarray}
The $(R,T)$ dependent part of the above equation decouples from the angular part (which is comprised of just the spherical harmonics) in the following way
\begin{eqnarray}
\left(\frac{\partial^2 \Phi_{\Omega}^{lm} }{\partial T^2}- \frac{\partial^2 \Phi_{\Omega}^{lm}}{\partial R^2} \right)+\frac{l(l+1)}{R(1-H^2R^2)}\Phi_{\Omega}^{lm} =0,
\label{kgeq}
\end{eqnarray}
where, we have used the properties of spherical harmonics $Y_{lm}(\theta,\phi)$. This equation shows that the $l\ne 0$ modes do not reach the Hubble scale simply because  {\em the effective potential becomes infinite there}. They are trapped inside the Hubble radius. If we, on the other hand, consider the $s-$waves  then they are solutions of the equation
\begin{eqnarray}
\left(\frac{\partial^2 \Phi_{\Omega}^{00} }{\partial T^2}-\frac{\partial^2 \Phi_{\Omega}^{00} }{\partial R^2}\right) =0.
\label{l=0}
\end{eqnarray}
This equation is valid for sub-huble modes $R\le 1/H$ since the metric \eqref{mRT2} is valid for this region. The mode solutions are
\begin{eqnarray}
\Phi_{\Omega}^{00} (T,R) \propto \exp{[-i\Omega (T \pm R)]}.
\end{eqnarray}
Now elevating $\Phi$ to the operator level, and introducing a new set of creation and annihilation operators ($b_{\Omega},~b_{\Omega}^{\dagger}$), and using a continuous basis of mode functions we express
\be
\Phi = \int{{d^3\Omega} (b_{\Omega > \Omega_H}^{L} U_{\Omega}^{\text{sub}} + b_{\Omega > \Omega_H}^{R} V_{\Omega}^{\text{sub}}}) + \textrm{h.c.},
\label{sub}
\ee
where $\Omega > \Omega_H$ ensures that the modes are sub-Hubble and {\it ``L''}/{\it ``R''} once again indicate the leftmoving/rightmoving modes with respect to an observer inside the Hubble radius. The mode functions are then easily found to be 
\ber
U_{\Omega}^{\text{sub}} &=&  \frac{1}{4\pi \sqrt{\Omega} R} ~e^{-i\Omega (T-R)} \label{uo1}\\
V_{\Omega}^{\text{sub}} &=&  \frac{1}{4\pi \sqrt{\Omega} R}~ e^{-i\Omega (T+R)}.\label{vo1}
\eer
 The vacuum states for these two sectors are then defined as $b_{\Omega >\Omega_H}^{in} | 0_{\text{T}}\rangle^{L} =0;~b_{\Omega >\Omega_H}^{R} | 0_{\text{T}}\rangle^{R}=0$. 

%%%%%%%%%%%%%%%%%%%%%%%%%%%%%%%%%%%%%%%%%%%%

Similarly, we can solve the wave equation, for super-Hubble modes in region-II, in the background of the metric \eqref{mRT3}. For that we use the ansatz
\begin{eqnarray}
\Phi=\displaystyle{\sum_{l,m,\Omega}}\frac{\Phi_{\Omega}^{lm}(T, R)}{T}Y_{lm}(\theta,u).
\end{eqnarray}
Under $s-$wave approximation the field operator is expanded as
\be
\Phi = \int{\frac{d^3\Omega}{4\pi \sqrt{\Omega}} ({b}_{\Omega <\Omega_H}^{L} U_{\Omega}^{\text{sup}} + {b}_{\Omega <\Omega_H}^{R} V_{\Omega}^{\text{sup}}}) + \textrm{h.c.}
\label{sup}
\ee
and mode functions for $T > R_H = 1/H$ are
\ber
U_{\Omega}^{\text{sup}} &=&  \frac{1}{T} ~e^{-i\Omega (T-R)} \label{uo2}\\
V_{\Omega}^{\text{sup}}&=&  \frac{1}{T}~ e^{-i\Omega (T+R)}. \label{vo2}
\eer

%%%%%%%%%%%%%%%%%%%%%%%%%%%%%%%%%%%%%%%%%%%
%\subsubsection{Completeness}

It is important now to comment on the completeness of various modes. In fact, from the above analysis, we see that for s-waves, it is possible to obtain a complete set of orthogonal modes (which can be normalised in discrete basis), by considering both the sub-Hubble and super-Hubble region, given in \eqref{sub} and \eqref{sup}. As a matter of fact, modes in the background of \eqref{mRT2} has a vanishing support in the super-Hubble region, whereas, for \eqref{mRT3} it is valid in the sub-Hubble  region. 

Therefore, it is clear that there will be two sets of Bogolyubov coefficients, pertaining the sub and super Hubble modes, while discussing particle creation for static observers in $(T,R)$ frame. This is formally different than the simpler two dimensional example discussed before.

%%%%%%%%%%%%%%%%%%%%%%%%%%%%%%%%%%%%%%%%%%%

\subsubsection{Bogolyubov coefficients and particle content}
We are interested in a specific case where the static observers in $(T,R)$ spacetime will look into the cosmological vacuum state $|0_{\text{C}}\rangle^{L/R}$ (``$L$'' stands for the left moving and ``$R$'' stands for the right moving modes). From our previous encounter with Unruh effect one expects $~^{L/R}\langle 0_{\text{C}}|b_\Omega^{\dagger L/R}b_\Omega^{L/R}|0_{\text{C}}\rangle^{L/R} \ne 0$, and leading us to the nontrivial case of particle creation. We want to calculate Bogolyubov coefficients for both left-moving and right-moving sectors and the particle content for these modes. 

Thus the most relevant Bogolyubov coefficient is $\beta_{\Omega\omega}$, and this gives the average number of particles in frequency $\Omega$ by the relation
\ber
\langle n_{\Omega}\rangle^{L/R} &=& ~^{L/R}\langle 0_{\eta}|b_\Omega^{\dagger L/R}b_\Omega^{L/R}|0_{\eta}\rangle^{L/R} \nonumber \\
                                                                   & =& \int_{0}^{\infty} d\omega |\beta_{\Omega\omega}^{L/R}|^2 \label{no}
\eer

Once again, by looking at the trajectories of static observers in $(T,R)$ frame in Fig. \ref{fig-1} we expect the particle creation to take place in the regions where the observer is accelerating or decelerating. We also expect, given the symmetry of the trajectories,  the particle content to be symmetric.

Here, we are interested in an observer inside the Hubble radius. The relevant mode functions, using $(T,R)$ coordinates, are calculated in subsection \eqref{subh} which are sub-Hubble modes.  Considering the left moving modes (with $V=const.$) the Bogolyubov coefficient of our interest is
\be
\beta_{\Omega\omega}^{L} = 2i \int_{V=const.} dU R^2 d\tilde{\Omega} U_{\Omega} \partial_{U} u_{\omega}.\label{bog}
\ee  
where $d\tilde{\Omega}= \sin\theta d\theta d\phi$. By choosing $V = 0$ hypersurface for integration we have
%\bwt
\be
\beta_{\Omega\omega}^{L} = -\frac{i}{2\pi \sqrt{\Omega\omega}} \int_{0}^{\infty} dU e^{-i(\Omega U + \omega\sqrt{2U/{\cal H}e})}\left(\frac{1}{U} + \frac{i \omega}{\sqrt{2{\cal H}eU}} \right).\label{int1}
\ee
%\ewt
For simplicity, we perform the integration by restricting ourselves to the situation such that the ratio between frequencies in two frames satisfy the condition $\frac{\omega}{\Omega}<<{\cal H}e$ (cosmological value of ${\cal H}\sim 10^{37} s^{-1}$). The physical implication of this is already discussed before for the two dimensional case. The final expression is found to be{\footnote{Detailed calculation is showed in Appendix \ref{outbog}.}
\be
\beta_{\Omega\omega}^{L} = -\frac{ie^{\frac{i\omega^2}{2{\cal H}e\Omega}}}{2\pi\sqrt{\Omega\omega}} \left(1-\omega\sqrt{\frac{\pi}{4{\cal H}e \Omega}}  + i\omega\sqrt{\frac{\pi}{4{\cal H}e \Omega}}\right)
\label{betaout}
\ee 
Now substituting $|\beta_{\Omega\omega}^{L}|^2$ in \eqref{no} and integrating we find the particle excitation number density as
\be
\langle n_{\Omega} \rangle^{L} = \frac{1}{16\pi^3 \Omega} \left(\log{\omega} -\sqrt{\frac{\pi}{{\cal H}e\Omega}}\omega + \frac{\pi}{4{\cal H}e \Omega} \omega^2 \right).  \label{nout}
\ee
The logarithmic term shows an infra-red divergence as $\omega\rightarrow 0$ in $\langle n_{\Omega} \rangle^{L}$ just like the case of particle creation by moving mirror \cite{birrel}. In comparatively higher frequency $\omega^2$ term dominates and if we substitute $\frac{\omega}{\Omega} = \epsilon e {\cal H}$ (with $\epsilon<<1$), we get
\be
\langle n_{\Omega} \rangle^{L} \simeq \frac{{\cal H}e}{64\pi^2}\epsilon^2
\ee
which is again quite small and interestingly independent of $\Omega$, just like the two dimensional case. It is, however, capable to contribute a finite energy density if one counts all the particles created during the entire lifetime of radiation stage of the universe, as well as, considers all the frequencies within the above mentioned limit. Again, it is necessary to employ a fully numerical analysis to go beyond this limit and we do not consider this in the present article. 

For the right-moving modes (with $U=const.$), the Bogolyubov coefficient is
\be
\beta_{\Omega\omega}^{R} = 2i \int_{U=const.} dV R^2 d\tilde{\Omega} V_{\Omega} \partial_{V} v_{\omega}
\ee  
We chose $U=0$ hypersurface for performing the integration over $V$ where $0 < V < \infty$. After simplifying, the above expression becomes
%\bwt
\be
\beta_{\Omega\omega}^{R} =  -\frac{i}{2\pi\sqrt{\Omega\omega}} \int_{0}^{\infty} dVe^{-i(\Omega V + \omega\sqrt{2V})}\left(\frac{1}{V} + \frac{i \omega}{\sqrt{2{\cal H}eV}} \right).
\label{boin}
\ee
%\ewt
which is identical to \eqref{int1}, with $U$ replaced by $V$ and, therefore, leads to $\langle n_{\Omega} \rangle^{L} = \langle n_{\Omega} \rangle^{R}$ in \eqref{nout}.

%%%%%%%%%%%%%%%%%%%%%%%%%%%%%%%%%%%%%%%%%%%%%%
%\subsubsection{Super-Hubble modes ($\Omega < \Omega_{H}$)}
Now let us consider the super-Hubble modes. While calculating the Bogolyubov coefficients in region II (i.e., super-Hubble region) we should consider the mode functions those have non-vanishing support in that region. The mode functions in $(\eta,~r)$ coordinates have non-vanishing support both at the sub-Hubble and super-Hubble regions. However, as we have already shown, in $(T,~R)$ coordinates, it is only the mode functions \eqref{uo2} and \eqref{vo2}, that have non-vanishing support in the super-Hubble region.  Therefore they will be used to calculate the Bogolyubov coefficients in region II. 

For the left-moving sector we need to calculate
\be
\beta_{\Omega\omega}^{'~L} = 2 i\int_{V} dU T^2 d\tilde{\Omega} U_{\Omega} \partial_{U} u_{\omega}.
\ee
where the relevant mode functions are given in \eqref{mod1} and \eqref{uo2}. Note that now the radius of the two sphere is given by $T$ which, in region II, is expressed in \eqref{ter}. After simplifying, and again integrating over the $V =0$ axis, the integral becomes
%\bwt
\be
\beta_{\Omega\omega}^{'~L} = -\frac{i}{2\pi\sqrt{\Omega\omega}} \int_{0}^{-\infty} dU e^{-i( - \Omega U + \omega\sqrt{-2U/{\cal H}e})}\left(\frac{1}{-U} + \frac{i \omega}{\sqrt{-2{\cal H}eU}} \right).\label{ints1}
\ee
%\ewt
Notice that, in region II, we have $-\infty < U <0$ and the above integration is identical to \eqref{int1} upon the replacement of $U$ by $-U$. Therefore, we end up with the result \eqref{nout} for an average particle excitation for the super-Hubble modes for an observer in region II. It is also trivial to check that one would end up with an identical expression for particle excitation for the right-moving modes in region II. This fulfils our expectation that, particle content for sub and super-Hubble region is identical, simply because of the symmetry of the trajectories in these two regions, shown in Fig. \ref{fig-1}.

This terminates our discussion on the particle creation in radiation dominated universe which is of course not an all encompassing picture, and in fact, it can be further extended in future works. Now, we shall move to the next section where another aspect of particle creation will be discussed via a different mechanism than the one we have seen so far - instead of an interchange between observers' frames, we shall see how the expansion of the universe itself excites  an initially defined vacuum state. 
%%%%%%%%%%%%%%%%%%%%%%%%%%%%%%%%%%%%%%%%%%%%%%

\section{Particle creation in multi-stage universe}

The need to go beyond the de Sitter case is obvious for the expanding universe since we need to consider other expansion stages following the inflationary universe. There are three more stages, such as the radiation, matter and dark energy dominated expansions, to consider if we want a unified view of cosmological expansion. For that we not only need to study each of the stages (epochs) individually but also to connect the physical variables while universe makes a transition from one epoch to another. It is a very important but challenging task to do in a fully realistic manner - such as by considering realistic values of lifetimes, going beyond the test field limit by considering the backreaction problem in a effective manner etc. Nonetheless, relaxing some of these considerations could still give a valuable insight and understanding of our cosmic history. In this part of the article, we shall review some earlier works  which deal some of these issues and set a platform for future studies.
%%%%%%%%%%%%%%%%%%%%%%%%%%%%%%%%%%%%%%%%%%%%%%%%%%%%%%%%
\subsection{Preliminaries}

Even before discussing particle creation in gravitational framework, we knew that quantum vacuum becomes unstable and gives rise to particle production when the external sources are sufficiently strong such as for strong electric fields (Schwinger effect) \cite{schwinger}.  When the external sources can be switched off asymptotically, leading to free field theory at very early and late times, one can define \textit{in} and \textit{out} vacua in the asymptotic regimes and study particle creation in a reasonably unambiguous manner.  However, many interesting cases  in the study of quantum field theory in curved spacetime do not allow us the luxury of asymptotic vacua. Cosmological particle creation is one such example. In this context, one can certainly calculate the time evolution of a given quantum state in Heisenberg or Schr\"odinger picture in an unambiguous manner. But interpreting the \textit{particle content} of this quantum state at any given time is fraught with ambiguities.  It is generally recognized that one cannot resolve these ambiguities by any unique procedure which is applicable in  all contexts. The best one could  do is to introduce different constructs which could probe different aspects of physics in the expanding background and develop an intuitive feel for the various phenomena. 

Another question, closely related to particle content, is the notion of classicality. A quantum field in an external background might have features which possess nearly classical description in certain contexts. For example, it is believed that fluctuations of  a scalar field, which were purely quantum mechanical in origin in the early stages of the inflationary phase, allow a description as purely classical stochastic fluctuation at the late stages. Much of the current paradigm in cosmology \cite{Brandenberger:1984cz, book-pad, rev-muk, rev-mar,  lyth2008, othr, othr1, othr2, othr3, othr4, othr5, othr6, othr7} pre-supposes such a notion of quantum to classical transition. Once again, it is not possible to quantify the degree of classicality of a field in a unique and all encompassing manner. The best we could do is to come up with measures of classicality and see how best they work in different contexts.

It would be useful, intuitively clear and somewhat economical if we could come up with constructs which simultaneously give a handle on the degree of classicality of the field and its particle content. In fact, such a criterion will sharpen our intuitive idea that well defined notion of particles will exist if and only if the degree of classicality is high, while the notion of particles will be drowned in the sea of quantum fluctuations when the degree of classicality is low. Such a procedure and a fairly comprehensive methodology was proposed sometime back in a series of papers \cite{gaurang2007, gaurang20071, gaurang2008}.
This work exploits the fact that the quantum theory of a minimally coupled scalar field in a Friedmann spacetime can be reduced to the study of a time dependent harmonic oscillator. 

The Schr\"odinger equation for the time dependent harmonic oscillator admits \textit{form-invariant} Gaussian states as solutions (this is well-known in the literature, see for example~\cite{sriram}), which --- in turn --- allows us to define {\it instantaneous} particle content of the state. Particles are defined in terms of instantaneous eigenstates specified at each moment. This approach has proved to be quite successful in dealing with time dependent particle content in various cases, like for example--- massless minimally coupled scalar field in de Sitter and radiation dominated Friedmann spacetime \cite{gaurang2007, gaurang20071}; complex scalar field in a constant electric field \cite{gaurang2007} and time dependent electric field \cite{gaurang2008} backgrounds. The quantum to classical transition is usually discussed in terms of Wigner function which links the wavefunction that appears in Schr\"odinger's equation to a probability distribution in phase space. However, it was found \cite{gaurang2007, gaurang20071} that the peaking of the Wigner function on the classical phase space trajectory is independent of particle content in several contexts and hence makes the definition of classicality in terms of Wigner function somewhat less useful. To counter this, an additional measure of phase space correlation (called the \textit{classicality parameter}) was proposed in \cite{gaurang2007, gaurang20071} which maintains our intuitive link between degree of classicality and particle content. When there is no particle creation this parameter is zero and in the presence of strong particle creation its modulus saturates at the maximum value of unity.
 
Motivated by these studies, we analysed the particle content and classicality of a quantum field, by considering the entire evolutionary history of our universe \cite{Singh:2013bsa}. Because the matter dominated phase in our universe has lasted only for about 4 decades of expansion, while the radiation dominated phase has lasted for nearly 24 decades of expansion, we approximate the evolutionary history of the universe as made up of just three stages --- the early (inflationary) de Sitter phase,  a radiation dominated phase and  late-time de Sitter phase characterizing the current accelerated expansion of the universe. The time-dependent minimally coupled scalar field equations are solved, separately, in these three stages. The field solutions as well as the scale factors corresponding different regions are then matched at two transition points (i.e., de Sitter to radiation and radiation to de Sitter transitions). These allow us to discuss issues like time dependent particle content, emergence of classicality for comoving case in an integrated manner.

\subsection{Schr\"{o}dinger dynamics of a Quantum Field}
\label{sec:schrdyn}

To set the stage, we will begin by summarizing the  formalism developed in~\cite{gaurang2007, gaurang20071} to study quantum fields in an expanding universe and recall the key ideas related to definition of states, particle creation and classicality. We will not provide detailed motivation of these ideas here; the interested reader may find more details in~\cite{gaurang2007, gaurang20071}. 

It is well-known that the dynamics of free fields in the Friedmann background can be reduced to that of decoupled, time dependent, harmonic oscillators which can be quantized in Schr\"odinger picture. We consider a massless minimally coupled scalar field in the spatially flat Friedmann background  
\ber
\label{metric}
ds^2 &=& dt^2 - a^2(t)d\tbf x^2 \nn\\
&=& a^2(\eta)(d\eta^2  - d\tbf x^2)
\eer
where the conformal time is defined by $\eta\equiv\int \dd t \, a^{-1}$. The action for the field is then given by
\ber
\label{actn1}
\calS[\Phi(\eta,\tbf x)] &=& \int \dd^4 x\,\sqrt{-g}\,\pp_a\Phi \pp^a\Phi \nn\\
&=&\f{1}{2}\int \dd^3 \tbf x\,\int\dd\eta\, a^2(\eta)\left(\pp^2_\eta\Phi - \pp^2_{\tbf x}\Phi\right).
\eer
Due to the translational invariance of the metric in \eq{metric}, one can  decompose the field into independent Fourier modes as 
\be
\Phi(\eta,\tbf x) = \int \f{\dd^3 k}{(2\pi)^3}\,\xi_{\tbf k}(\eta) e^{i\tbf{k}\cdot\tbf{x}}
\ee
Since $\Phi$ is real, this implies $\xi_{\tbf k} = \xi_{-\tbf k}^*$ for the  complex $\xi_{\tbf k}$. This constraint essentially halves the degrees of freedom in $\xi_{\tbf k}$, so that in terms of a single real variable $\phi_{\tbf k}$, one can express the action in \eq{actn1} as
\be
\label{reducedactn}
\calS[\phi_{\tbf k}(\eta)] =\f{1}{2}\int \dd^3 \tbf k\,\int\dd\eta\, a^2(\eta)\left(\dot{\phi}_{\tbf k}^2 - k^2\phi_{\tbf k}^2\right)
\ee
where the dot implies the derivative with respect to $\eta$ and $k = |\tbf k|$. The field system thus gets reduced to a bunch of decoupled harmonic oscillators in the Fourier domain with time-dependent mass, $a^2(\eta)$ and frequencies, $k$. We can now now use the fact that this Schr\"{o}dingier equation admits  time-dependent, form-invariant, Gaussian states with vanishing mean given by:
\ber
\label{wavefn}
\psi(\phi_{\tbf k},\eta) &=& N \exp\left[-\alpha_k(\eta)\phi_{\tbf k}^2\right] \nn\\
&=& N \exp\left[-\f{a^2(\eta) k}{2}\left(\f{1- z_k}{1+z_k}\right)\phi_{\tbf k}^2\right]
\eer
The time evolution of the  wave function is now given in terms of that of the functions $\alpha_k(\eta)$ and $z_k(\eta)$ which satisfy the equations:
\be
\label{eqnforalpha}
\dot{\alpha}_k = \f{2\alpha_k^2}{a^2} - \f{1}{2} a^2 k
\ee
and 
\be
\dot{z}_k + 2ikz_k + \left(\f{\dot{a}}{a}\right)(z_k^2 -1) = 0
\ee
Here we use the notation and terminology introduced in \cite{gaurang2007} in which $z_k$, called the {\it excitation parameter},  measures the deviation of $\alpha_k$ from the adiabatic value. (The functions $\alpha_k$ and $z_k$ depend only on the modulus of \tbf k and hence the subscripts are not in boldface.) We thus need to solve for $\alpha_k$ or $z_k$ to infer the quantum evolution of the system and related characteristics. The non-linear first order equations can be related to the second-order \textit{linear} differential equation by introducing  another function $\mu_k(\eta)$, defined through the relation $\alpha_k = - (ia^2/2) (\dot{\mu}_k/\mu_k)$, which satisfies:   
\be
\label{eqnformu}
\ddot{\mu}_k + 2\left(\f{\dot{a}}{a}\right)\dot{\mu}_k + k^2 \mu_k = 0
\ee
This is the same as the field equation for $\phi_\tbf k$ resulting from varying the action in \eq{reducedactn}. As for the function $z_k$, it is related to $\mu_k$ by:
\be
\label{zmu}
z_k = \left(\f{k\mu_k +  i \dot{\mu}_k}{k\mu_k -  i \dot{\mu}_k}\right)
\ee
Thus it suffices to solve for $\mu_k$ given the boundary conditions to determine the quantum evolution of the system. 

Since $\mu_k$ satisfies the second order linear differential equation, it will have two linearly independent solutions and thus we can write, in general, $\mu_k(\eta) = \mathcal{A}_k s_k(\eta) + \mathcal{B}_k  s_k^*(\eta)$. But  $z_k$ and $\alpha_k$ depends only on  the ratio $\dot{\mu}_k/{\mu_k}$ so that the overall normalization of $\mu_k$ is irrelevant and the evolution only depends on the ration $\mathcal{R}_k = \mathcal{B}_k/\mathcal{A}_k$ for given initial conditions. 

The initial conditions are set such that the state described by the wave function in \eq{wavefn} is a ground state with zero particle content at some time say $\eta = \eta_i$ when $a (\eta_i) = a_i $. (It is often convenient to use the scale factor itself as a  time variable with the  replacements $d/d\eta \rightarrow \dot{a}\,d/da$ etc.) Then, the initial condition of the wave function at $a=a_i$ being the ground state wave function of an harmonic oscillator demands,
\be
\alpha_k(a_i) = \f{a^2_i k}{2}
\ee
or equivalently $z_k(a_i) = 0$,  implying
\be
\label{cndtnmu}
\left(\f{\dot{a}}{\mu_k}\left.\f{\dd{\mu}_k}{\dd a}\right)\right|_{a_i} = i k
\ee
which in turn determines $\mathcal{R}_k$ thereby fixing the state. As the system evolves, we are interested in the two specific quantities: the particle content of the state at any time and the degree of classicality of the state. Both of these were discussed in detail in \cite{gaurang2007, gaurang20071} and we will just summarize the motivation here and adopt the ideas: Our initial condition implies that the state begins as a ground state at $a=a_i$ but at any later time will be different from the instantaneous ground state. To quantify the instantaneous particle content of this state it is then reasonable to compare it with the \emph{instantaneous} eigenstates at every instant obtained by adiabatically evolving the  the eigenstates at some initial epoch. Since $\psi$ is an even function, the overlap is non-zero only with even eigenstates. One can calculate this overlap (for details the reader is referred to \cite{gaurang2007, gaurang20071}) with the eigenstates to get time-dependent probability distribution of transitions and using it, the mean number of quanta in the state at any time can be computed.
This particle content is given in terms of $z_k$ as
\be
\label{n}
\bra n_k\ket = \f{|z_k|^2}{1-|z_k|^2}
\ee 
Note that being time-dependent and related to transitions within the instantaneous eigenstates, we do \textit{not} expect $\bra n_k\ket$ to be monotonic in general. The mean occupation number can go up and down and hence should not be taken to be the `particle' content in the \emph{classical} sense since it can be accompanied by fluctuating \emph{quantum} noise when the system is far away from classicality. 

The degree of classicality of the state brings us to the next quantity of interest, the \emph{classicality} parameter $\mathcal{C}_k$, which is the measure of phase space correlations of the system. Classicality is usually  quantified by the use of Wigner distributions and  is inferred from the peaking of the distribution on the corresponding classical trajectory. However, it can be shown \cite{gaurang2007, gaurang20071} that the naive reliance on Wigner function can lead to ambiguities. Hence a new correlation function, $\mathcal{C}_k$ was introduced in \cite{gaurang2007, gaurang20071} as a more robust measure to quantify classicality and it is found to be in excellent agreement with our intuitive ideas in many cases. This quantity is given by
\be
\mathcal{C}_k = \f{\mathcal{J}_k\sigma_k^2}{\sqrt{1+(\mathcal{J}_k\sigma_k^2)^2}}
\ee  
where $\mathcal{J}_k$ and $\sigma_k$ are the parameters of Wigner function defined in the $\phi_\tbf k - \pi_\tbf k$ phase space of the oscillator for the Gaussian state in \eq{wavefn}
\be
\mathcal{W}(\phi_{\tbf k},\pi_{\tbf k},\eta) = \f{1}{\pi}\exp\left[-\f{\phi^2_{\tbf k}}{\sigma_k^2} - \sigma_k^2(\pi_\tbf k - \mathcal{J}_k \phi_\tbf k)^2\right].
\ee
In terms of $z_k$ we have,
\be
\label{cp}
\mathcal{C}_k = \f{2\,\mathrm{Im}(z_k)}{1 - |z_k|^2}.
\ee
The vanishing of $\mathcal{C}_k$ implies $\mathcal{J}_k = 0$ and Wigner distribution is then an uncorrelated product of gaussians in $\phi_\tbf k$ and $\pi_\tbf k$ which is the case for the ground state which itself is gaussian, set up as the initial condition. Otherwise the classicality parameter is confined to the interval $[-1,1]$ when the Wigner function becomes correlated. The particle creation and classicality of a state are strongly associated with each other; as we shall see the notion of particles become well-defined when the degree of classicality is high and vice-versa. With the structure in place, we shall get on with our study of these aspects in the cosmological context.

%%%%%%%%%%%%%%%%%%%%%%%%%%%%%%%%%%%%%%%%%%%%%%%%%%%%%%%%

\subsection{The Three Stage Universe}
\label{sec:3phases}

We consider a three stage universe consisting of an initial inflationary de Sitter phase characterized by the hubble parameter, $H_\inf$ which evolves into a radiation dominated phase and ends up in a late-time de Sitter phase dominated by a cosmological constant $\Lambda$. Equivalently the final phase can be characterized by the hubble parameter, $H_\Lambda$ with $H_\Lambda^2=\Lambda/3$. (This model ignores the matter dominated phase for simplicity, which can be justified by the fact that --- in our universe ---  the matter domination lasts only for about 4 decades while radiation domination lasted for about 24 decades.) The scale factor for such a three-stage universe can be taken to be:
\begin{figure}[t!]
\includegraphics[scale=0.50,width=0.49\textwidth]{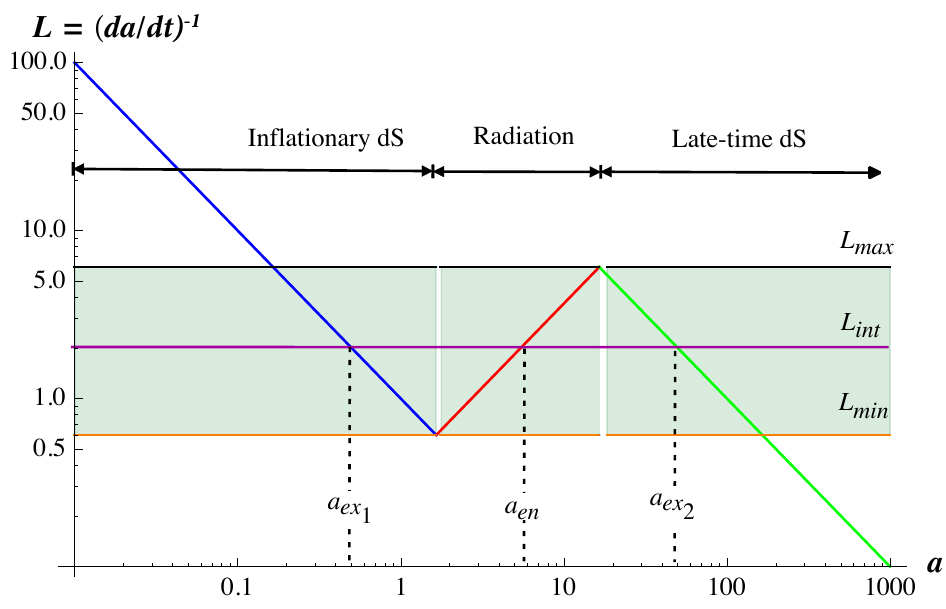}
\caption{\emph{Our three stage universe}. Evolution of comoving hubble radius, $L = (da/dt)^{-1}$ with the scale factor $a$ for $\epsilon = 0.01$. in the inflationary phase  I(blue), radiation dominated phase (red) and in late-time  de Sitter phase II (green) the lines have  slopes $\pm 1$ in the logarithmic plot. We  have two characteristic length scales, $L_{max}$ and $L_{min}$ forming a band such that any length scale within the band has three transition points where it goes super-hubble, sub-hubble and finally super-hubble again.}
\label{comovuniv}
\end{figure} 

\begin{multline}
\label{aoft}
a(t) =
\left
\{
\begin{array}{ll}
 e^{H_\inf t}  & \hspace{10pt}t \le t_r \\
        (2H_\inf e)^{1/2}\, t^{1/2} &  \hspace{10pt} t_r\le t \le t_{\Lambda}\\
        (H_\inf/H_\Lambda)^{1/2}e^{H_\Lambda t}  &\hspace{10pt}  t \ge t_{\Lambda}
\end{array}
\right.
\end{multline}
where $t_r = (2H_\inf)^{-1}$ and $t_{\Lambda} = (2H_\Lambda)^{-1}$ are the comoving times at the two respective transitions and we ensured that the scale factor and its derivative are continuous at the transition points. Further, we shall take  $H_\Lambda = \epsilon H_\inf = \epsilon H$ with $\epsilon\ll 1$, since for our real universe, $\epsilon\sim10^{-54}$. However,  for visual clarity of figures, we will use sufficiently small  values for $\epsilon$ to discuss the effects and comment about the cosmological case later. 
We can re-express the scale factor in terms of the conformal time  as 
\begin{multline}
\label{aofeta}
a(\eta) =
\left
\{
\begin{array}{ll}
 \left[(\eta_r - \eta)H+ e^{-1/2}\right]^{-1}  & \hspace{10pt}\eta \le \eta_r \\
 (e/\epsilon)^{1/2}+ He (\eta - \eta_\Lambda) &  \hspace{10pt} \eta_r \le \eta \le \eta_\Lambda\\
  \left[(\epsilon/e)^{1/2} - \epsilon H (\eta - \eta_\Lambda)\right]^{-1} & \hspace{10pt}  \eta \ge \eta_\Lambda
        \end{array}
\right.
\end{multline}
with $(\eta_f - \eta_\Lambda) = (He^{1/2})^{-1}(\epsilon^{-1/2} -1)$. But it is more convenient to use  $a$ itself as the time-variable since it gives a better conceptual understanding of the length scales involved. Note that $a_r = a (t_r) = e^{1/2}$ and $a_\Lambda = a(t_\Lambda) = (e/\epsilon)^{1/2}$ at the transition points with our choice of normalization. The comoving hubble radius, in terms of the scale factor, is given by
\begin{multline}
\label{comovlength}
   L(a) = (\dot{a})^{-1} = \left\{
     \begin{array}{ll}
        (H a)^{-1} &\hspace{7pt}a \le e^{1/2} \\
        (a/ H e) &  \hspace{7pt} e^{1/2}\le a \le (e/\epsilon)^{1/2}\\
        (a\epsilon H)^{-1}  &\hspace{7pt}   a\ge (e/\epsilon)^{1/2}
     \end{array}
   \right.
\end{multline}  
The above scheme is pictorially depicted in \fig{comovuniv}. In the logarithmic plot the lines have unit slope and are at $45$ degrees. The comoving hubble radius shrinks during the initial inflationary de Sitter (blue) phase till $a_r = e^{1/2}$, when the radiation dominated phase starts and the comoving hubble radius increases (red) until $a_\Lambda = (e/\epsilon)^{1/2}$ (which is decided solely by the value of $\epsilon$) the final de Sitter phase (green) sets in leading to the shrinkage again.

In such a universe, there exist two length scales, $L_{max} = 1/(H\epsilon^{1/2} e^{1/2})$ and $L_{min} = 1/(H e^{1/2})$ and a band in between which is special (see e.g. \cite{cosmin, cosmin1}). A wave mode characterized by the length scale larger than $L_{max}$, once exits the comoving hubble radius in the initial de Sitter phase, will remain super-hubble at all times. Similarly modes with wavelengths smaller than $L_{min}$ remains sub-hubble till it exits the hubble radius in the late-time de Sitter phase. Any wave mode characterized by the length scale, $L_{int}$ lying within the band encounters three transition points: it exits the hubble radius at some $a_{ex_1}$ during initial de Sitter phase and goes super-hubble, then enters the the hubble radius at $a_{en}$ during the radiation dominated phase becoming sub-hubble again and re-exits the hubble radius in the final de Sitter phase at $a_{ex_2}$ becoming super-hubble once again. While $L_{min}$ is independent of $\epsilon$, $L_{max}$ depends on it inversely which is expected, since $\epsilon$ determines the duration of the radiation dominated phase. This rich terrain, as we shall see, affects the particle creation aspects and classicality for a test scalar field in a non-trivial manner. 

We shall now apply the formalism of \sec{sec:schrdyn} in the toy cosmological model described above. This requires working out the evolution of the wave function in the three stages which, in turn, requires computing $\mu_k(a)$, with a given  initial condition in the first de Sitter phase and sewing it with the other two patches by demanding the continuity of the wave function and its derivative at the transition points (i.e., de Sitter $\rightarrow$ radiation $\rightarrow$ de Sitter transitions). We shall now turn to this task and describe the solution in the three stages.
 
%%%%%%%%%%%%%%%%%%%%%%%%%%%%%%%%%%%%%%%%%%%%%%%%%%%%%%%%
\subsubsection{The inflationary phase}
\label{subsec:ds1}
With the scale factor specified for the case in \eq{aoft} or \eq{aofeta}, we can solve for $\mu_k(\eta)$ using \eq{eqnformu} and by inverting the function $a(\eta)$ we have,
\be
\mu_k^{(1)} (a) =  s_k (a) + \mathcal{R}_k s_k^*(a)
\ee
with
\be
s_k(a) =  \left(\f{1}{a} - \f{i H}{k}\right) \exp\left[\f{ik}{H}\left(1-\f{1}{a}\right)\right].
\ee
The constant $\mathcal{R}_k$ is determined by imposing the initial condition as per \eq{cndtnmu} at an initial epoch $a_i$, which gives:
\be
\mathcal{R}_k (a_i) = \left(1-\f{2ik}{Ha_i}\right)^{-1}\exp\left[\f{2ik}{H}\left(1 -  \f{1}{a_i}\right)\right]
\ee
Different choices of $a_i$ will correspond to different initial conditions. We will choose the state to be a ground state in the asymptotic past  that is, when $a_i\rightarrow 0$.  This can be achieved by choosing  $\mathcal{R}_k=0$  for all $k$. Thus we have
\be
\label{muk1}
\mu_k^{(1)} (a) =   \left(\f{1}{a} - \f{i H}{k}\right) \exp\left[\f{ik}{H}-\f{i k}{a H}\right]
\ee
This is the state evolved to an epoch $a$ from conventional Bunch-Davies vacuum ~\cite{bunchdavies} at the asymptotic past. At any finite time it is different from the instantaneous vacuum state and is a mixture of positive and negative frequency modes~\cite{suprit1302} with non-zero particle content. 
 
%%%%%%%%%%%%%%%%%%%%%%%%%%%%%%%%%%%%%%%%%%%%%%%%%%%%%%%%
\subsubsection{Radiation dominated phase}
\label{subsec:radtn}
For the radiation dominated phase, $\mu_k(a)$ is given by
\be
\label{muk2}
\mu_k^{(2)} (a) =  \f{1}{a}\left(C_k e^{-ik a/eH } + D_k e^{ik a/eH}\right)
\ee
where $C_k$ and $D_k$ are determined by the matching conditions at $a =  a_r  = e^{1/2}$ 
\be
\label{match12}
\mu_k^{(1)} (a_r) = \mu_k^{(2)} (a_r);\hspace{10pt} \mu_k^{(1)'} (a_r) = \mu_k^{(2)'} (a_r)
\ee
where the prime denotes derivative with respect to $a$. This can be done analytically but the resulting expressions are not very illuminating.

%%%%%%%%%%%%%%%%%%%%%%%%%%%%%%%%%%%%%%%%%%%%%%%%%%%%%%%%
\subsubsection{The late-time de Sitter phase}
\label{subsec:ds2}

The scale factor in the late-time de Sitter is related to the inflationary phase by the replacements $H\rightarrow \epsilon H$ and an overall scaling by $\epsilon^{-1/2}$. So we have 
\be
\label{muk3}
\mu_k^{(3)} (a) =  E_k \bar{s}_k (a) + F_k \bar{s}_k^*(a)
\ee
with
\be
\bar{s}_k(a) = \exp\left[\f{ik}{\epsilon H}\left(1-\f{1}{a}\right)\right] \left(\f{1}{a} - \f{i \epsilon H}{k}\right).
\ee
Again, $E_k$ and $F_k$ are determined by matching at $a =  a_\Lambda  = (e/\epsilon)^{1/2}$: 
\be
\label{match23}
\mu_k^{(2)} (a_\Lambda) = \mu_k^{(3)} (a_\Lambda);\hspace{10pt} \mu_k^{(2)'} (a_\Lambda) = \mu_k^{(3)'} (a_\Lambda).
\ee
With the help of \eq{match12} and \eq{match23} we now have $\mu_k(a)$ connected up for all the three stages. The next step is to compute the particle content and the classicality parameter in an integrated manner for the entire evolution history of the universe. This is done using \textit{Mathematica} for algebraic manipulations and the results are presented in the next section.  

%%%%%%%%%%%%%%%%%%%%%%%%%%%%%%%%%%%%%%%%%%%%%%%%%%%%%%%%

\subsection{Particle content and Classicality}
\label{sec:pcc}
We now recall the method summarized in \sec{sec:schrdyn} to find the average particle number and classicality parameter given by \eq{n} and \eq{cp} respectively. The exact analytical results for these quantities are simple in the inflationary  phase, but gets algebraically unwieldy for the radiation phase and an impossibly complicated (having hundreds of terms!) in the final de Sitter phase. Hence we shall first present our results in \fig{epsilon1} and \fig{epsilon2} using results for a range of values of $\epsilon$ using algebraic manipulation software. Having described the exact results in this manner, we will provide approximate analytic expressions highlighting the behavior pattern in various limits in \sec{subsec:analyticlimits}.

%%%%%%%%%%%%%%%%%%%%%%%%%%%%%%%%%%%%%%%%%%%%%%%%%%%%%%%%

\subsubsection{The numerical results}
As is evident from \eq{aofeta}, the smaller values of $\epsilon$ correspond to longer lifetime of the radiation dominated phase. Here we first consider toy universes with $\epsilon = 0.01$ and $\epsilon = 0.0001$ to capture the whole picture  which is numerically  difficult to do with the cosmologically relevant value of $\epsilon$. Note that, the sub and super-hubble regions, as shown in \fig{comovuniv}, also provide an estimate for $k/H$. Modes with $k/H < e^{1/2} \epsilon^{1/2}$ are always super-hubble once they exit the hubble radius in the inflationary phase and modes with $k/H > e^{1/2}$ are always sub-hubble until they exit the hubble radius in the late-time de Sitter phase. Any other mode will exit, enter and re-exit the hubble radius in different stages as shown in \fig{comovuniv}. For the set of expressions we come across here,  it is possible to normalize the the wave vector ($k$) with respect to $H$ so that one can set $H = 1$ for the numerical scheme. But it should be remembered that in this case $k$ really means $k/H$. We shall reinstate $H$ in the next subsection where we deal with analytical expressions. 

The excitation parameter $z_{k}$ is calculated separately for three stages from \eq{zmu}. Since we have three different expressions for $\mu_{k}$, given by \eq{muk1}, \eq{muk2} and \eq{muk3} in different stages there exist three corresponding results for $z_{k}$ viz., $z^{(1)}_{k},~z^{(2)}_{k}$ and $z^{(3)}_{k}$ which by construction match at the transition points. Then it is straight forward to calculate $\langle n_k \rangle$ and ${\cal C}_{k}$ separately for each stage and plot them collectively as in \fig{epsilon1} and \fig{epsilon2}.  

\begin{figure*}[t!]
\includegraphics[width=0.40\textwidth]{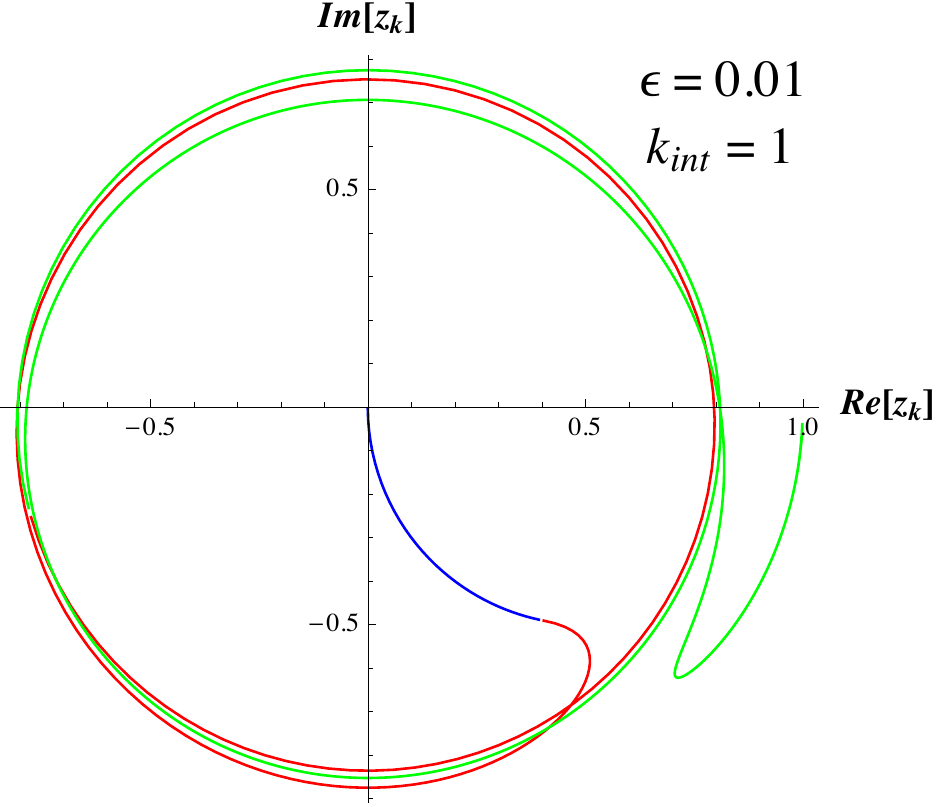}\hfill
\includegraphics[width=0.49\textwidth]{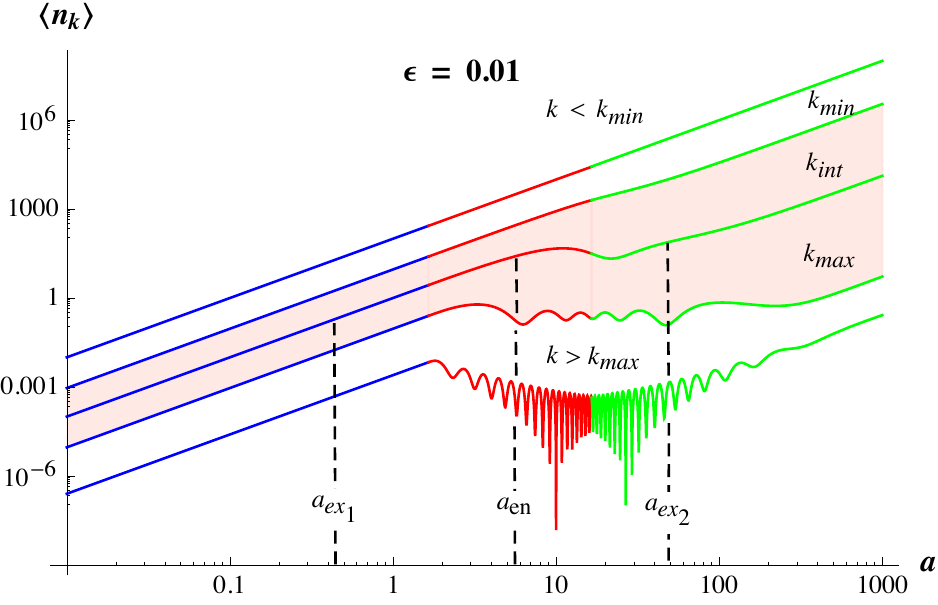}
\includegraphics[width=0.33\textwidth]{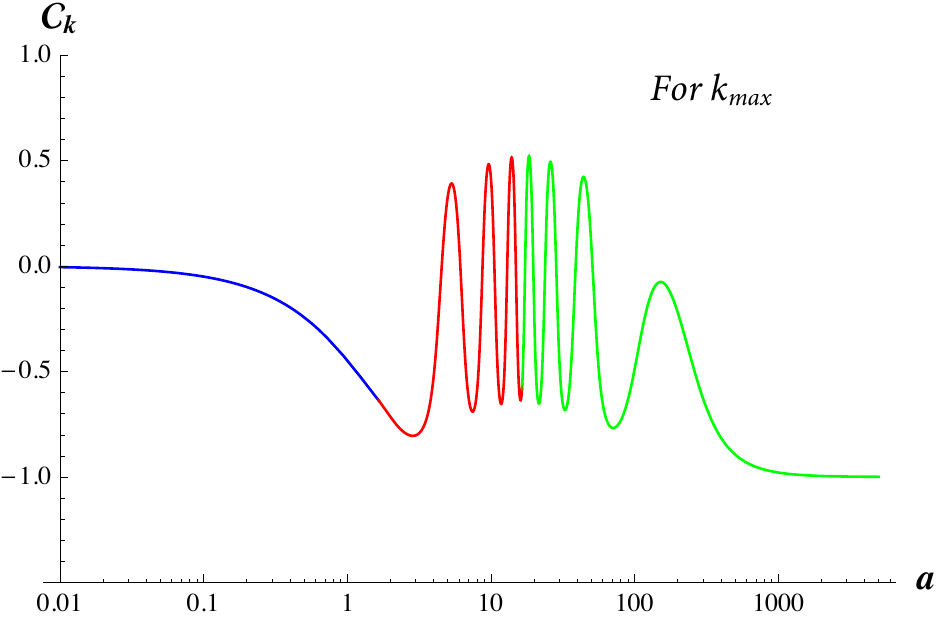}\hfill
\includegraphics[width=0.33\textwidth]{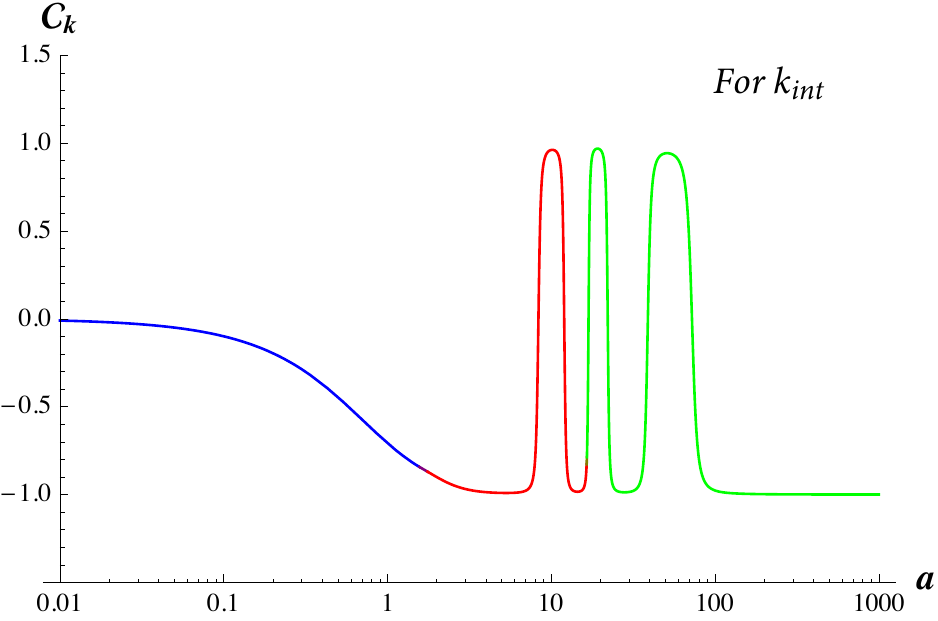}\hfill
\includegraphics[width=0.33\textwidth]{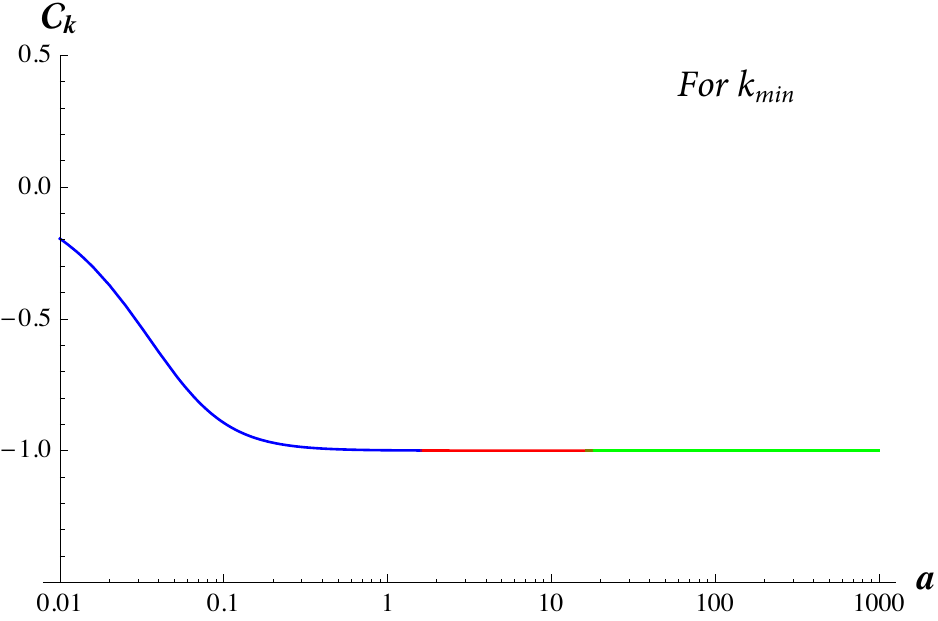}\hfill
\caption{Evolution  of the excitation parameter $z_{k}$, average particle number $\langle n_{k} \rangle$ and classicality parameter ${\cal C}_{k}$ with the scale factor for $\epsilon = 0.01$. The color scheme is as follows: blue $\rightarrow$ early inflationary phase, red $\rightarrow$ radiation dominated phase, green $\rightarrow$ late-time de Sitter phase. The modulus of the wave vector $k$ is fixed in each plot. Modes which have $k < k_{min}$ are always super-hubble and those with $k > k_{max}$ are always sub-hubble until they finally exit the late-time de Sitter phase. Any other intermediate mode ($k_{int}$) exit the Hubble radius in early de Sitter phase at $a_{ex_{1}}$ then enters in the radiation dominated phase at $a_{en}$ and finally re-exits in late-time de Sitter phase at $a_{ex_{2}}$. Here $k_{max} =1.64872 $, $k_{min} = 0.164872$, $k_{int} = 1$ for $z_k$ and ${\cal C}_k$ and $k_{int} = 0.5$ for $\langle n_{k} \rangle$ while $k = 10,~0.05 $ for lower and uppermost plots respectively, outside the band.}
\label{epsilon1}
\end{figure*} 

\fig{epsilon1} depicts the case with $\epsilon=0.01$. The excitation parameter starts from $z_{k}^{(1)}=0$ (shown in blue) in the inflationary phase and it gradually shifts toward $z_{k}^{(3)}=1$ (shown in green) in the late-time de Sitter. As we have already mentioned, the non-zero values of the excitation parameter ($z_{k}$) implies that the system has departed from its initial adiabatic vacuum state given by $z_{k}^{(1)}=0$. The appearance of the radiation dominated phase in between makes $z_{k}^{(2)}$ (shown in red) to follow a circular trajectory for a while and --- somewhat interestingly --- even after this phase has ended, $z_{k}^{(3)}$ (green) continues to exhibit the same behavior, before it changes its direction to reach unity. This behavior of $z_k$ is particularly interesting because for a single de Sitter case $z_{k}$ does not follow a circular path \cite{gaurang2007, gaurang20071} which is clearly a `residual memory' of the radiation dominated phase. That is, despite the fact that the background spacetime has already made a transition from radiation to late-time de Sitter phase, the scalar field does not `know' about this for a while until $z_{k}^{(3)}$ leaves the circular trajectory. This is a  signature of non-adiabatic behavior that we mentioned above.

The above characteristics of $z_{k}$ have nontrivial consequences on $\langle n_{k} \rangle$ and ${\cal C}_{k}$. It is known that $\langle n_{k} \rangle$ follows a power law (and hence appears as a straight line in logarithmic plot) for pure de Sitter phase as previously found in \cite{gaurang2007, gaurang20071}. This is because the average number of particles that are being created in this phase increases monotonously with the expansion until it reaches a mode dependent maximum value at the point of transition. This holds for $a \le e^{1/2}$ since the initial de Sitter phase (blue) has no prior information about the future transition. After the first transition, once the universe is in the radiation dominated phase (red), the behavior of $\langle n_{k} \rangle$ changes dramatically. This change basically depends upon the modulus of the wave vector ($k$). For super-hubble modes ($k < k_{min}$), the power-law behaviour of $\langle n_{k} \rangle$ is unaffected because these modes have exited the hubble radius in the inflationary phase itself. The $\langle n_{k} \rangle$ of  modes corresponding to $k > k_{min}$ starts {\it oscillating} once the universe changes to radiation dominated phase. This amplitude of the oscillation increases with $k$ and is most pronounced for sub-hubble modes. One can directly relate this oscillatory behavior of $\langle n_{k} \rangle$ with the circular trajectories of $z_{k}$. Just as in the case of $z_k$, for $\langle n_{k} \rangle$ also, we see the effect of radiation dominated phase is `remembered' in the late-time de Sitter phase (green); in fact, this effect persists for quite some time. Note that in a pure de Sitter universe $\langle n_{k} \rangle$ has no oscillations. Therefore appearance of oscillations in the late-time de Sitter phase is a residual effect of the radiation dominated era. However, for large $a$, after the relevant modes become super-hubble, this oscillatory behavior in $\langle n_{k} \rangle$ settles down to power-law behaviour. One important observation that follows from \fig{epsilon1} and \fig{epsilon2} is that irrespective of the value of $\epsilon$ the oscillation in $\langle n_{k} \rangle$ approaches a saturation value toward the end of the radiation phase. In fact, in the absence of the late-time de Sitter phase, $\langle n_{k} \rangle$ remains to be fixed at this saturated value forever (as it happens for a pure radiation-dominated universe \cite{gaurang2007, gaurang20071}). The latter phase makes $\langle n_{k} \rangle$ to shift from its saturated value it would have reached in the radiation dominated phase. However, as we pointed out before, because of the `memory' it is not possible to instantaneously drive the system away from this saturation value. Some amount of time has to be spent in the late-time de Sitter phase for this to occur, which varies from mode to mode and turns out to be larger for sub-hubble ($k > k_{max}$) regime. 

Interestingly, this saturation in average particle number has a similarity with the particle creation in electric field that was encountered earlier in \cite{gaurang2007, gaurang20071, gaurang2008}. In the presence of constant electric field, at late times, $\langle n_{k} \rangle$ becomes nearly constant \cite{gaurang2007, gaurang20071}. For time dependent electric field there are two cases which were discussed in \cite{gaurang2008}:  For smaller values of a dimensionless parameter $\sqrt{qE}t$ (where $q$ is the charge and $E$ is the electric field), the asymptotic mean particle number depends upon the duration for which the field ($E$) was nonzero, whereas,  for considerably larger values of $\sqrt{qE}t$, the final particle content becomes independent of this duration. Similar results hold in our case. The asymptotic average particle number near the end of radiation phase fluctuates for smaller lifetime of the radiation phase (for example, with $\epsilon=0.01$ in  \fig{epsilon1}) and for relatively larger lifetime it becomes constant (like in  \fig{epsilon2} with $\epsilon = 0.0001$). We shall return to this discussion once again in the next subsection where we deal with analytical results.     

The classicality parameter ${\cal C}_k$ starts from zero in the beginning of inflationary phase and depending upon $k$ and $\epsilon$ it shows different characteristics as depicted in \fig{epsilon1} and \fig{epsilon2}. Any mode with $k = k_{int}$ which lies within the $[k_{min},\,k_{max}]$ band, tends to a classical description near the end of the inflationary phase as ${\cal C}_k \rightarrow -1$ but as the universe makes a transition to radiation phase, it starts oscillating. These oscillations last  during the radiation phase as well as in the beginning of the late-time de Sitter phase. In the late-time de Sitter phase when a mode exits the hubble radius at large $a$ one finds  ${\cal C}_k$ saturates at -1. This property of the mode establishes a connection between the classicality and its hubble exit. After the hubble  exit, in the early and late-time de Sitter  ${\cal C}_k \rightarrow -1$ and modes behave classically (as expected). But in between, when the mode is sub-hubble, it oscillates and remains away from classical description. To clearly understand this aspect we have plotted ${\cal C}_{k}$ by considering two modes which are sub-hubble  (for the first two phases) and is super-hubble  (once it exits from the inflationary phase). The sub-hubble mode does not reach -1 in the first two phases but once it exits the hubble radius in the late-time de Sitter phase it reaches that value. On the other hand, the super-hubble mode, once it exits from the initial de Sitter phase remains always super-hubble and saturates with ${\cal C}_k \rightarrow -1$. This relation of classicality and hubble exit is, of course, known (\cite{Brandenberger:1984cz, book-pad, rev-muk, rev-mar,  lyth2008, othr}) in the context of primordial perturbations but our procedure provides a quantitative measure of degree of classicality. 

To understand the oscillatory nature in the average particle number in \fig{epsilon1} and \fig{epsilon2} one should again refer to the behavior of classicality parameter. The fluctuations in ${\cal C}_k$ imply that the system is in the quantum domain and the notion of average number of particles at a particular instant is not well defined in the `classical' sense. At most one can ask, in loose sense, a time averaged value of  $\langle n_{k} \rangle$ and interpret this as the number of particles produced during that time interval. The particle content is well-defined and has the intuitive behaviour of monotonic increase only when degree of classicality is high which is precisely what we would expect. While our formalism allows us to define $\langle n_{k} \rangle$ at any time, one cannot really think of them as `particles' when its value is oscillatory. This is precisely what happens  when degree of classicality is low (as is to be expected) and particle definition becomes ambiguous. As we stressed right at the beginning of the paper we do not want to over emphasize any given notion of particle in a strong field regime; in stead we want to correlate the behaviour of a well-defined parameter (measured by $\langle n_{k} \rangle$) with degree of classicality. this study confirms our intuitive expectations and adds strength to the interpretation of both our definition of particle content and degree of classicality. 
\begin{figure*}[t!]
\includegraphics[width=0.40\textwidth,scale=0.2]{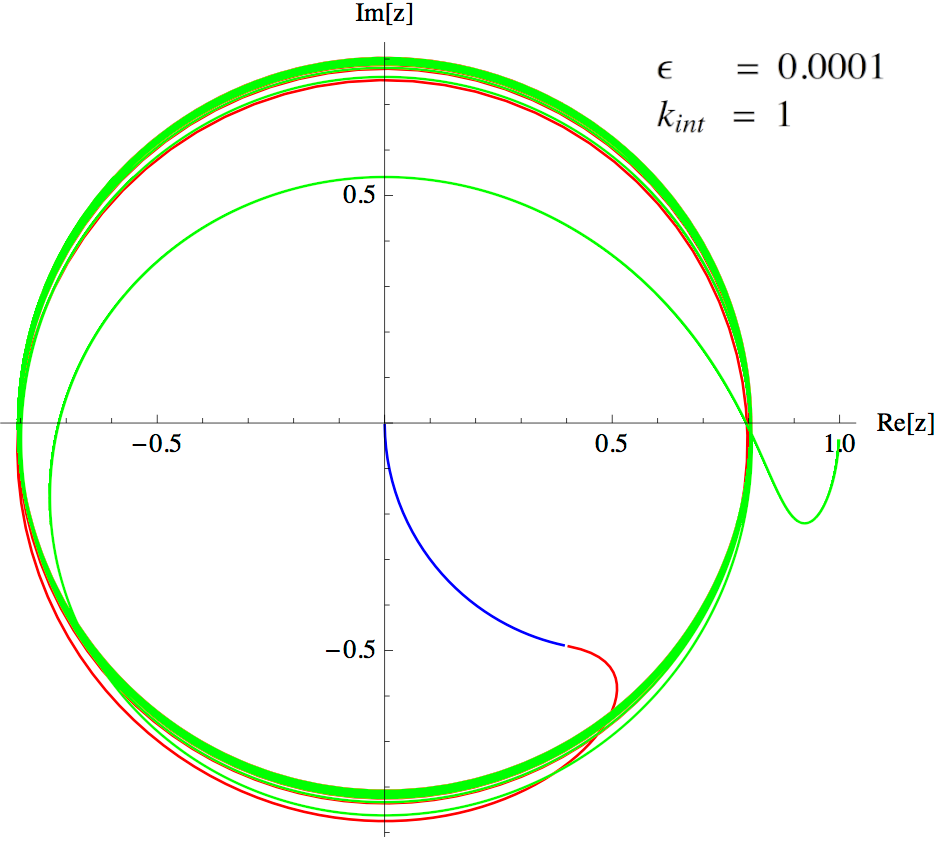}\hfill
\includegraphics[width=0.49\textwidth]{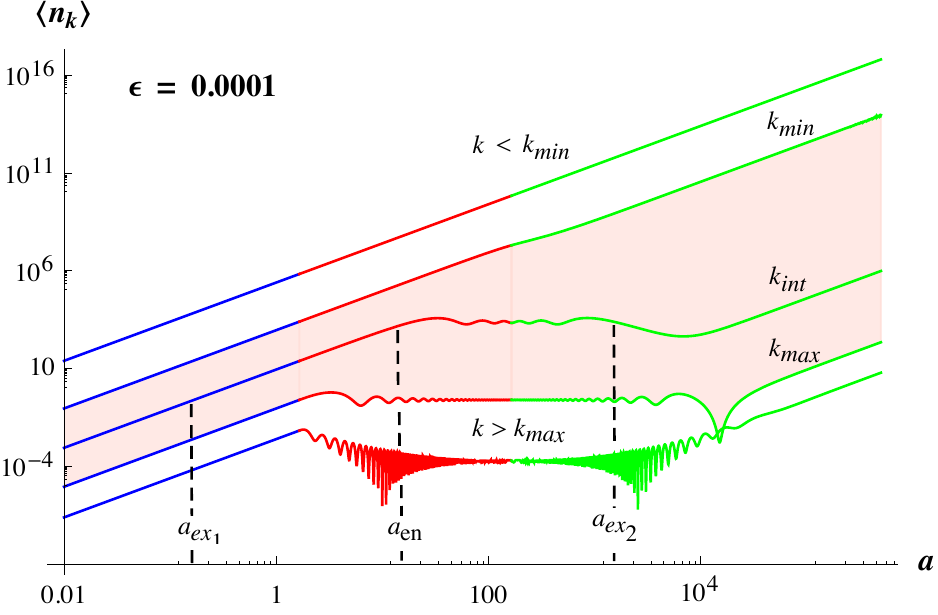}
\includegraphics[width=0.33\textwidth]{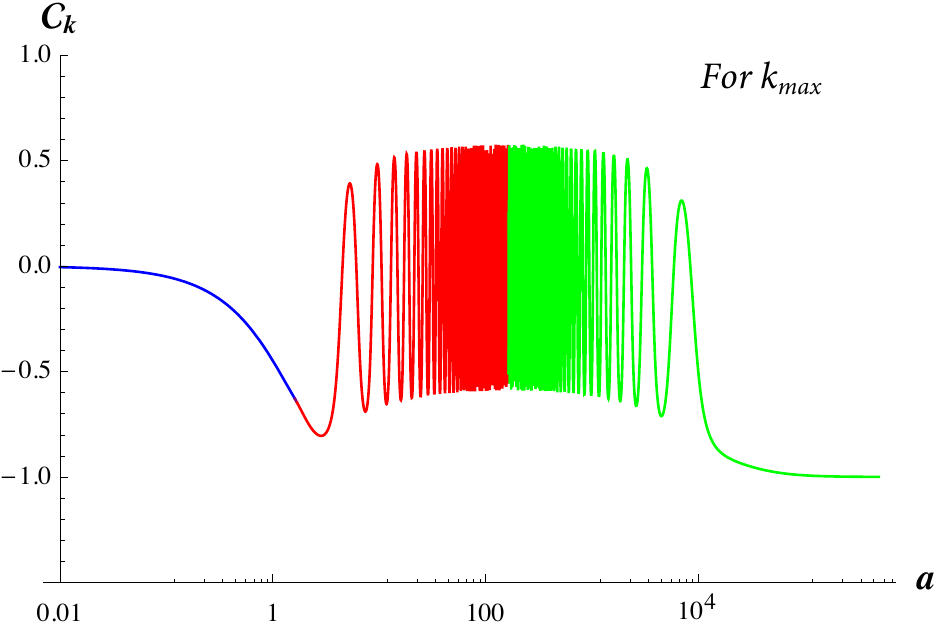}\hfill
\includegraphics[width=0.33\textwidth]{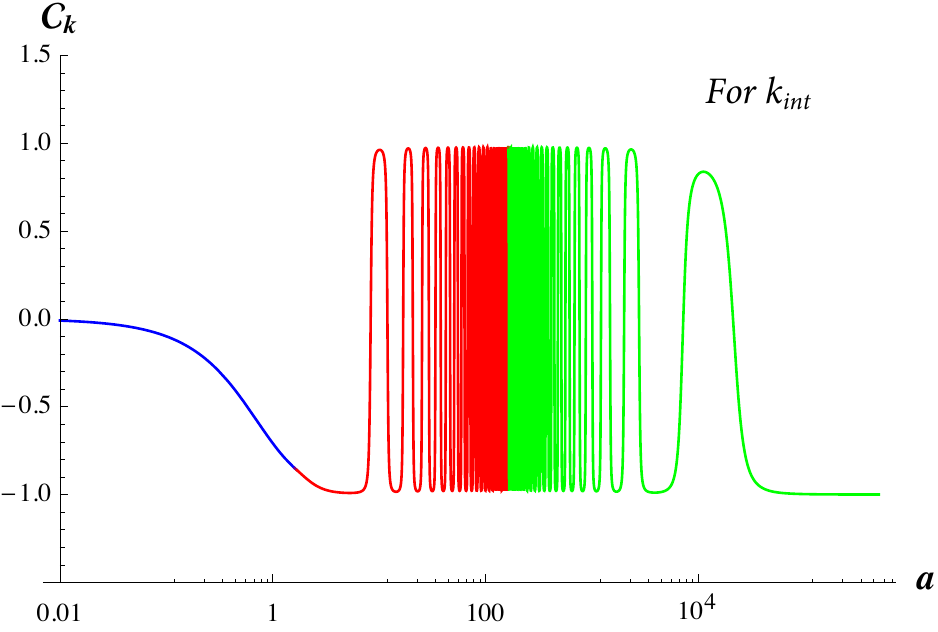}\hfill
\includegraphics[width=0.33\textwidth]{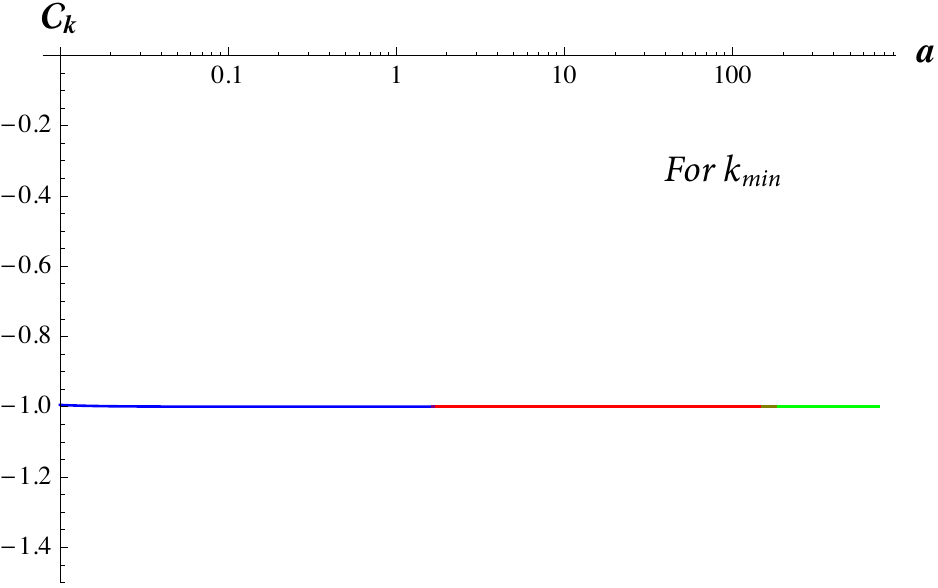}
\caption{Evolution of the excitation parameter $z_{k}$, average particle number $\langle n_{k} \rangle$ and classicality parameter ${\cal C}_{k}$ with the scale factor for $\epsilon = 0.0001$. Color scheme and notations are same as \fig{epsilon1} but in this case $k_{max} = 1.6487$, $k_{min} = 0.01648$ and $k_{int} = 0.167$. Outside the band, $k = 10.0$ (lower) and $k = 0.001$ (upper). For $z_{k}$ and $\mathcal{C}_{k}$ we have taken $k_{int} = 1.0$.}  
\label{epsilon2}
\end{figure*} 

Almost all physical features of \fig{epsilon1} remain intact for $\epsilon = 0.0001$ as is evident from \fig{epsilon2}. The only difference arises in terms of number of rotations encountered in the plot of $z_{k}$. Since for this case the radiation phase has a comparatively longer lifetime the oscillations of $\langle n_{k} \rangle$ and ${\cal C}_k$ persist for a longer duration. In fact, as we make $\epsilon$ smaller, this winding number increases substantially and becomes unmanageably high for cosmological value of $\epsilon \sim 10^{-54}$. We shall now look at this case analytically, taking suitable limits, in the next subsection.

%%%%%%%%%%%%%%%%%%%%%%%%%%%%%%%%%%%%%%%%%%%%%%%%%%%%%%%%
\subsubsection{Analytic limits in different regimes}
\label{subsec:analyticlimits}
As we have already mentioned, the analytical results become increasingly complex as we proceed with the evolution of the universe and thus require suitable approximations for the final late-time de Sitter phase. For the inflationary phase, the expressions for various quantities are simple, and are given by
\be
\label{z1}
z_k^{(1)} = \frac{a H}{a H+2 i k}
\ee

\ber
\label{n1}
  \langle n_{k}^{(1)} \rangle &=& \frac{a^2 H^2}{4 k^2} \\
  {\cal C}_{k}^{(1)} &=& -\frac{a H}{k \sqrt{\frac{a^2 H^2}{k^2}+1}}
\eer
\\
These match with the results found earlier in \cite{gaurang2007}. For any value of $k/H$, in the logarithmic plot, $\langle n_{k}^{(1)} \rangle\propto a^2$ is a straight line  as shown in \fig{epsilon1} and \fig{epsilon2} (blue lines). Also, as expected, none of the expressions above depend upon $\epsilon$. From the expression of the classicality parameter it is obvious that super-hubble modes with $aH/k >> 1$ have ${\cal C}_{k}^{(1)} \approx -1$ and behave classically. However, all other modes remain away from classical description. In this phase, we can also calculate the \textit{comoving} energy density by multiplying \eq{n1} by $k$ (since $\omega_k = k$) and integrating over $\mathrm{d}^3k$ as
\ber
\mathcal{E}^{(1)} (a) &=& \f{a^2H^2}{8\pi^2  }\int_{k_{a}}^{k_{b}} k \dd k\nn\\
&=& \f{a^2H^2}{16\pi^2}(k_{b}^2 - k_{a}^2).
\eer 
The above expression has a UV divergence which is usual in the case in quantum field theory and requires a cutoff. The origin and cure for divergences in de Sitter phase is still a matter of debate in the literature but for our purpose we shall just introduce a cut-off. We can choose $k_a = k_{min}$ and $k_b = k_{max}$ for the energy in the band $[k_{min}$,\,$k_{max}]$. For the cosmological case, this band is wide enough to be interesting with $k_{min} = H(e\epsilon)^{1/2}$ and $k_{max}= H e^{1/2}$. The background comoving energy density is $\mathcal{E}^H_{bg} = 3H^2/8\pi G$. So the ratio of the energy densities at the end of inflation is,
\be
\label{comp1}
\f{\mathcal{E}^{(1)} (a_r)}{\mathcal{E}^{H}_{bg}} = \f{e^2H^2 G}{6\pi}(1 -\epsilon) \sim L_p^2 H^2 
\ee
where $L_p$ is the planck length and we have $\epsilon<<1$. With $E_p = 10^{19}$ GeV and $E_{GUT} = 10^{15}$ GEV, this ratio is of order $10^{-16}$ which shows that $\mathcal{E}_{\mathrm{inf}}<<\mathcal{E}_B$. As a result  the particle creation in the band we are studying does not substantially backreact on the background geometry. 

In the radiation dominated phase, the exact analytical results for $z_k^{(2)}$ and $\langle n_{k}^{(2)} \rangle$ are a little cumbersome and are given by
\bwt
\ber
\label{z2}
z_k^{(2)} &=& \frac{e H \left(e^{\frac{2 i k}{\sqrt{e} H}} H (e H+2 i a k)-e^{\frac{2 i a k}{e H}} \left(e H^2+2 i \sqrt{e} H k-2 k^2\right)\right)}{-e^{2+\frac{2 i k}{\sqrt{e} H}} H^3+e^{\frac{2 i a k}{e H}} (e H-2 i a k) \left(e H^2+2 i \sqrt{e} H k-2 k^2\right)} \\ \label{n2}
  \langle n_{k}^{(2)} \rangle &=& \frac{e^2 H^2}{8 a^2 k^6} \left(e^2 H^4+2 \left(a^2 H^2 k^2+k^4\right) -\left(e^2 H^4+4 a \sqrt{e} H^2 k^2-2 e H^2 k^2\right) \text{Cos}\left[\frac{2 \left(-a+\sqrt{e}\right) k}{e H}\right] \right. \notag \\
 && \left.  -2 kH \left(\left(-a+\sqrt{e}\right) e H^2+2 a k^2\right) \text{Sin}\left[\frac{2 \left(-a+\sqrt{e}\right) k}{e H}\right]\right). \eer
\ewt 
Note that although the scale factor in a radiation dominated phase has an explicit $\epsilon$ dependence in \eq{aofeta}, both $z_{k}^{(2)}$ and $\langle n_{k}^{(2)} \rangle$ are not explicit functions of $\epsilon$. Let us now examine the behavior in the special cases which are of physical interest. For super-hubble modes ($k < k_{min} =  (\epsilon e)^{1/2}$), it turns out that 
\be
\langle n_{k}^{(2)}\rangle  \approx \frac{a^2 H^2}{4 k^2} + \mathcal{O}(k^2)
\ee
which exactly matches with $\langle n_{k}^{(1)} \rangle$ in \eq{n1}. Not surprisingly, for these modes the log plot is a straight line with slope 2 as shown in \fig{epsilon1} and \fig{epsilon2}. One can always change $\epsilon$ to compare with the cosmological case and in such a case $k_{min}$ is also shifted to a much lower value which is proportional to $\epsilon^{1/2}$. On the other hand for sub-hubble modes ($k > k_{max}$) the argument of the Sine and Cosine terms in \eq{n2} get bigger and the oscillatory nature takes over from monotonic behaviour. Asymptotically, as the universe approaches the end of radiation phase  (for $a \rightarrow e^{1/2}/\epsilon^{1/2}$), the $\langle n_{k}^{(2)} \rangle$ tends to a saturation value
\be 
\label{n2satu}
{\langle n_{k}^{(2)} \rangle}_{sat} \approx \frac{e^2 H^4}{4 k^4}
\ee
when we ignore the smaller oscillatory terms. This is approximately the average number of particles at the end of radiation phase and correspond to the saturated regime of the plots in \fig{epsilon1} and \fig{epsilon2}. Again, it is straightforward to find the average energy density due to these particles by multiplying (\ref{n2satu}) by $k^3dk/2\pi^2$ and integrating over all $k$ (from some $k_a$ to $k_b$) to give:
\be
\label{ensatu}
{\cal E}^{(2)}_{sat} = \frac{e^2H^4}{8\pi^2}\ln\Big|\frac{k_b}{k_a}\Big|.
\ee
The average energy density per logarithmic mode is a constant and is proportional to the fourth power of the inflationary hubble parameter. We can again compare its value with the comoving background energy density at the end of radiation phase which is
\be
\mathcal{E}^{\mathrm{rad}}_{bg} = \f{3H^2 e^2}{8\pi L_p^2}
\ee
to get
\be
\f{{\cal E}^{(2)}_{sat}}{\mathcal{E}^{\mathrm{rad}}_{bg} } = \f{L_p^2 H^2}{3\pi} \ln(\epsilon^{-1/2}) \sim L_p^2 H^2
\ee
which essentially remains almost the same as in \eq{comp1} since there is not much particle creation during the radiation phase due to saturation. The backreaction due to the modes in the band we are studying is not a concern for the evolution of background geometry. 

For the late-time de Sitter phase, we have given the exact expression for $z_k^{(3)}$ in the ~\app{app:z3}. (The exact analytic expression for ${\langle n_{k}^{(3)} \rangle}$ is too cumbersome to offer any insight and hence is not included.) Let us consider this expression in the appropriate limits. First, note that due to matching conditions ${\langle n_{k}^{(3)} \rangle}$ is equal to the value of ${\langle n_{k}^{(2)} \rangle}$ as given in \eq{n2satu} at the beginning of this phase for all $k$. Further, for any $k$, at late-times ${\langle n_{k}^{(3)} \rangle}$ varies as $a^2$ leading to straight lines of slope 2 in log-log plots. Again, for all super-hubble modes with $k<k_{min} = (\epsilon e)^{1/2}$ in the small $k$ limit, this  behavior continues with
\be
z_k^{(3)} \approx 1-\frac{2 k^2}{a^2 H^2}-\frac{2 i k}{a H \epsilon }
\ee
which then gives
\be
\langle n_{k}^{(3)}\rangle  \approx \frac{a^2 H^2}{4 k^2} 
\ee
with the functional form of ${\langle n_{k}^{(3)} \rangle}$ matching ${\langle n_{k}^{(1)} \rangle}$given in \eq{n1}. These features are remain valid for  the realistic cosmological value of $\epsilon$, for which the numerical value of $k_{min} \sim \epsilon^{1/2} \sim 10^{-27}$ is extremely small. 

The situation is more complicated for the field modes with $k>k_{min}$. Some of these modes fall in the intermediate band $k_{min}<k<k_{max}$ and others remain sub-hubble until they exit late-time de Sitter. To quantify the modes in the intermediate band is difficult analytically, but we can study the other limit of $k>k_{max}$. For these modes, with small $\epsilon$  and  large $a$ (since $a > (e/\epsilon)^{1/2}$) for the third region, the dominant terms of ${\langle n_{k}^{(3)} \rangle}$  are given by

\begin{figure}[t!]
\includegraphics[scale=0.5]{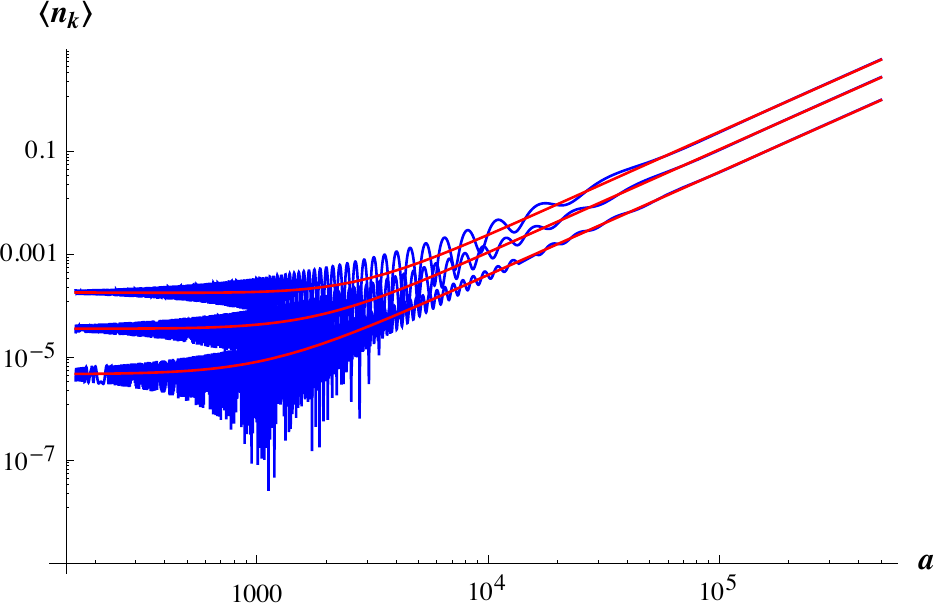}
\caption{Comparison of the average particle number $\langle n_{k}^{(3)}\rangle$ (shown in blue) with its average value ignoring the oscillations (shown in red) as given by the expression in~\eq{analyticn3} in the large $k$ limit. Here $\epsilon = 0.0001$ and $k/H = 10,15, 25$ increasing downwards.}
\label{analyticcompare}
\end{figure} 

\ber
\label{analyticn3}
{\langle n_{k}^{(3)} \rangle} &\approx& \frac{e^2 H^4}{4 k^4} - \frac{2H^4 e^{3/2} a\epsilon}{4 k^4}\nn\\
&&+\frac{H^2  \left(e^2H^4-e H^2k^2+k^4\right)a^2\epsilon^2}{4 k^6}
\eer
This  expression summarizes the behavior of particle content the third region when the oscillations are averaged out in and is shown in \fig{analyticcompare}. The zeroth order term is the saturation value of ${\langle n_{k}^{(2)} \rangle}$ which ${\langle n_{k}^{(3)} \rangle}$ picks up for relatively small $a$. Later on, when when $aH\epsilon/k >> 1$, the $a^2$ behaviour becomes the dominant feature; the behaviour shown in~\fig{epsilon1} and~\fig{epsilon2} corroborates this, i.e., when $a > k/H\epsilon$, the oscillations die down and monotonic behaviour arises. 

Finally, we plot the energy density of the field in the $[k_{min},\,k_{max}]$ band given by
\be
\label{energy}
\mathcal{E} = \f{1}{2\pi^2}\int_{k_{min}}^{k_{max}} \langle n_k\rangle k^3 \dd k
\ee
in \fig{energydensity} for $\epsilon = 0.0001$ and $H=1$. The energy density increases sharply (red) in the inflationary phase due to significant  particle creation in this phase and is then almost constant in the radiation phase (blue) when  particle creation is insignificant. The saturated remnant is also seen early in the third phase (green). The energy density of the field is quite low as compared with the comoving background energy density and does not pose any backreaction issues for the modes we have studied.   

\begin{figure}[t!]
\includegraphics[scale=0.6]{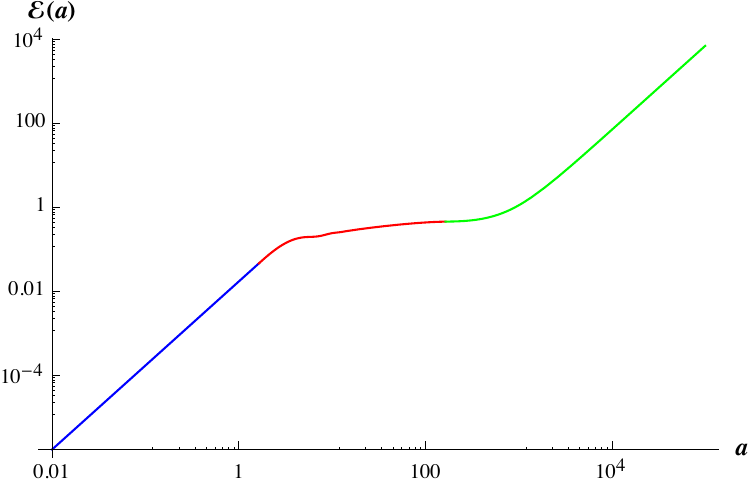}
\caption{Evolution of the energy density of the field in the $[k_{min}$,\,$k_{max}]$ band as given by \eq{energy} for $\epsilon = 0.0001$ and $H=1$. The color scheme is the same as in previous figures.}
\label{energydensity}
\end{figure}

%%%%%%%%%%%%%%%%%
\subsection{Implications for generation of perturbations from inflation}
\label{subsec:implications}

Finally we want to comment briefly on the connection between this work and the generation of perturbations in the inflationary scenario. In the standard approach to this problem, one computes the quantum fluctuations of the field at the time when the mode exits the Hubble radius in the initial de Sitter phase. This is usually done in the form of the two-point function $\langle 0|\phi(x)\phi(y)|0\rangle$. The Fourier transform of this  C-number is then identified with the stochastic fluctuations of a \textit{classical} random field at the time of re-entry of the mode to the Hubble radius (see e.g., p 637 of \cite{paddy2}). The usual justification for this procedure is based on two factors: (a) The modes behave classically once they are well outside the Hubble radius. (b) It is assumed that once they become classical they stay classical and hence can be described by standard perturbation theory after they re-enter the Hubble radius. 

Though this scenario is by now widely accepted (and the results of such a computation agrees well with observations), it must be noted that  the quantum to classical transition of the density perturbations is still not completely well understood from a conceptual point of view. 
We believe the approach introduced this paper will throw more light on this issue. 
The key point is that we now have an intuitively clear, quantitative, measure for the classicality of the fluctuations and we need not deal with a `two-level' description of the fluctuations being either fully classical or fully quantum mechanical. We will confine ourselves to brief comments here and hope to address this issue more comprehensively in a separate work. 

Our discussion of the field modes \textit{does} confirm the standard assumption (a) above, viz. that modes become classical when they leave the Hubble radius. What is more is that we can quantify the degree of classicality as the universe evolves. But our analysis also shows that, when the mode re-enters the Hubble radius the the degree of classicality does not stay constant but rapidly oscillates. This fact can have important implications for structure formation scenarios which are based on assumption (b) above, viz. that once the fluctuations are classical, they remain classical. But we must stress that, in the work reported here, we have treated the quantum field as purely a test field and did not incorporate the back reaction due to perturbations in the geometry. In the correct approach we need to take into account both the metric fluctuations (in particular the scalar mode representing the Newtonian potential in a particular gauge) and the field fluctuations as a coupled system (see e.g, p. 636 of \cite{paddy2}). It seems natural to treat the scalar mode perturbations of geometry as classical, in which case, we have the standard scenario of a quantum system interacting with a classical one and we need to consider issues like decoherence (which is often invoked to explain the classicality of perturbations though this may not be completely satisfactory). Further, we also note that the classicality parameter oscillates rapidly when the modes have re-entered the Hubble radius which is a different behaviour compared to the one exhibited when they were originally sub-Hubble radius in the inflationary phase. (This is clearly seen in the behaviour of $z_k$ in all the figures). This difference tells us that the two situations are not identical and our interpretation needs to take this difference into account. So the question cannot be answered only based on the results of the work described here and we need to extend it by taking into consideration the back reaction of geometry both in the quantum and classical regimes as well as the oscillations of the classicality parameter.

Similar comments apply to the study of the tensor components of the metric perturbations because they obey essentially the same type of wave equation as a scalar field in the cosmological context. Once again the conventional wisdom is that: (a) The gravity wave modes are quantum mechanical until they leave the Hubble radius in the inflationary phase but become classical when they are outside the Hubble radius.
(b) Once they become classical they remain classical as a stochastic gravity wave background today. The result in (a) can again be justified by an analysis similar to ours because the mathematics is essentially the same. But in tackling the issue of (b) it seems difficult to invoke effects due to back reaction [unlike in the case of scalar field modes] and it may be the oscillations of the classicality parameter which contains the relevant information. These are the issues needs to be addressed in future works.

%%%%%%%%%%%%%%%%%%%%%%%
\section{Discussions}
The aim of this article is to review few interesting aspects of particle creation in and beyond the de Sitter stage of expanding universe and their immense importance in our understanding of cosmology. We wanted to highlight the fact that putting relevant  coordinates and observers help us to sharpen our intuition about some physical phenomena since they give us a new framework to study them. For example, in radiation dominated universe, using the new ($(T,R)$) coordinates we found interesting new physics encountered by the fundamental cosmological observers (FCOs) - they will be exposed to a radiation, via particle creation, due to their own motion! This is very similar to the accelerated observers in Minkowski spacetime, with of course some differences which we also discussed in this review. We find this discussion particularly interesting because our understanding of cosmos relies on the FCOs and any new possibility of an observation (even in principle) can tell us something new about the radiation era of early universe which is by itself an interesting research ground. Although our discussion was complete for two dimensional toy model, there is a hope to extend its completeness  in four spacetime dimensions.

By extending the field modes beyond de Sitter universe we obtained a unified view of particle content for the massless scalar field modes in and after the inflationary stage, i.e., into the radiation and dark energy dominated universe. This gave an understanding, at the principal level by ignoring the backreaction problem, how the transition from one epoch to another affects the particle content of the field modes. We also discussed when the particle notion is well defined and when it is fraught with ambiguities, in terms of a ``classicality parameter''. The importance of the classical vs quantum nature of the particle definition was shown to have an indication of a deep connection on the problem of generation of classical density perturbation from the quantum fluctuation of the inflaton field. 

Going beyond the de Sitter stage is therefore quite interesting and holds deep insights that might be crucial for our unified view of cosmic history and this review article aims to bring this matter into the attention of the community.

%%%%%%%%%%%%%%%%%%%%%%%%%%%%%%%%%%%%%%%%%%%
\section{Acknowledgements}
A large part of this review is based on my collaborative work with T. Padmanabhan and Suprit Singh. I thank Satoshi Iso and Douglas Singleton for several illuminating discussions.
%%%%%%%%%%%%%%%%%%%%%%%%%%%%%%%%%%%%%%%%%%%

\appendix
\section{Derivation of the metric}
\label{newst}
Let us consider a conformal transformation, of power law type, on the original (null) cosmological FRW coordinates, given by
\begin{eqnarray}
V= \frac{1}{\lambda m} v^m \label{Vv}\\
U = \frac{1}{\lambda m} u^m\label{Uu}
\end{eqnarray}
for $u>0$ (with a constant $\lambda$) and,
\begin{eqnarray}
V= \frac{1}{\lambda m} v^m \label{Vv}\\
U = -\frac{1}{\lambda m} u^m\label{Uu}
\end{eqnarray}
for $u<0$, respectively. Using a power law expression for the scale factor  $a (t)=a_0 t^n$ we have $\eta = \int\frac{dt}{a(t)} = \frac{t^{1-n}}{a_0 (1-n)}$ and $a (\eta)=a_0 (a_0 (1-n)\eta)^{\frac{n}{1-n}}$. Using $\eta=\frac{u+v}{2}$ and $u,~v$ \eqref{Uu} and \eqref{Vv}, respectively, gives $a = a(U,V)$. Thus the metric \eqref{ncf} in new ($U,V$) null coordinates is,
\begin{eqnarray}
ds^2 = A (U,V) dU dV - B(U,V) (d\theta^2 + \sin^2\theta d\phi^2),
\label{newUV}
\end{eqnarray} 
where,
%\bwt
\begin{eqnarray}
A (U,V) &=&  (\lambda^2 a_0^2) (\pm \lambda^2  m^2 U V)^{\frac{1}{m}-1}  [a_0(1-n)/2]^{\frac{2n}{1-n}}    \left(  (\pm \lambda m U)^{1/m}+(\lambda  m V)^{1/m}\right)^{\frac{2n}{1-n}} \label{A} \\
B (U,V) &=& \frac{a_0^2}{4} [a_0(1-n)/2]^{\frac{2n}{1-n}} \left( (\lambda  m V)^{1/m} - (\pm \lambda m U)^{1/m}\right)^2 \left(  (\pm \lambda m U)^{1/m}+(\lambda  m V)^{1/m}\right)^{\frac{2n}{1-n}} \label{B}  \nonumber \\
\end{eqnarray}
%\ewt

The above metric takes a very interesting form for radiation dominated phase ($m=2$ and $n=1/2$) where \eqref{A} and \eqref{B} becomes 
\begin{eqnarray}
A (U,V) &=& \left(\frac{\lambda^2 a_0^4}{4}\right) \frac{(\sqrt{V} \pm \sqrt{\pm U})^2}{4 \sqrt{\pm UV}}\\
B (U,V) &=& \left(\frac{\lambda^2 a_0^4}{4}\right) \left(\frac{V- (\pm U)}{2}\right)^2.
\end{eqnarray}
Therefore, the metric \eqref{newUV} takes the form \eqref{newrd}. Note that in order to get \eqref{newrd} we have set $\frac{\lambda^2 a_0^4}{4} =1$. By equating the scale factor and its derivative at the transition point for a universe transiting to radiation stage  from the inflationary stage one can easily calculate $a_0=\sqrt{2{\cal H}e}$\cite{Singh:2013bsa}, implying
\be
\lambda = 1/{\cal H}e \label{lmda}
\ee
where ${\cal H}$ is the Hubble constant of the inflationary universe. Therefore the basic transformation between the FRW cosmological coordinates and new coordinates read \eqref{Uu-n}.

%%%%%%%%%%%%
\section{Calculation of Bogolyubov coefficient \eqref{nout}}
\label{outbog}
First using the mode function \eqref{uo1} and the relationship between the coordinates \eqref{Uu-n} we obtain
\be
\partial_U u_{\omega} = \frac{e^{-i\omega u}}{4\pi \sqrt{\omega}R}\left(\frac{1}{V-U} -  \frac{i\omega}{\sqrt{2{\cal H}e U}} \right)
\ee
Substituting $U_\Omega$ from \eqref{uo1} and the above expression, and integrating over the angular part in \eqref{bog} we get  \eqref{int1}. Next we substitute $U$ by $u$ using the relationship \eqref{Uu-n} in \eqref{int1}
\be
\beta_{\Omega\omega} = -\frac{i}{2\pi\sqrt{\Omega\omega}} \int_0^{\infty} du e^{-i\omega u - i\Omega {\cal H} e u^2/2}(i\omega + 2/u)
\ee

\be
\beta_{\Omega\omega} = -\frac{ie^{\omega^2/2{\cal H} e\Omega}}{2\pi\sqrt{\Omega\omega}} (i\omega I_{1} + 2I_{2}) \label{bogn}
\ee

\ber
I_{1} &=& \int_0^{\infty} du e^{\frac{-i\Omega{\cal H} e}{2}(u+\frac{\omega}{{\cal H} e\Omega})^2}\\
I_{2} &=& \int_0^{\infty} \frac{du}{u} e^{\frac{-i\Omega{\cal H} e}{2}(u+\frac{\omega}{{\cal H} e\Omega})^2}
\eer
Now we make a change of variable by setting $x=(u+\frac{\omega}{{\cal H} e\Omega})^2$. This imply
\ber
I_{1} &=& \int_{x_0}^{\infty}\frac{ dx }{2\sqrt{x}}e^{\frac{-i\Omega{\cal H} e}{2}x}\\
I_{2} &=&  \int_{x_0}^{\infty}\frac{ dx }{2\sqrt{x}(\sqrt{x} -\sqrt{x_0})}e^{\frac{-i\Omega{\cal H} e}{2}x}\eer
where $x_0 = (\frac{\omega}{{\cal H} e\Omega})^2$. In the limit $\omega/\Omega << {\cal H}$ we have $x_0 \rightarrow 0$ and using the identity \eqref{id} it is easy to check
\ber
\lim_{x_0\rightarrow 0} I_1 &=& e^{i\pi/4} \sqrt{\frac{\pi}{2{\cal H}e \Omega}}\\
\lim_{x_0\rightarrow 0} I_2 &=& 1/2.
\eer
Finally substituting these expressions in \eqref{bogn} we find \eqref{nout}.
%%%%%%%%%%%%%%%%%%%%%%%%%%

%%%%%%%%%%%%%%%%%%%%%%%%%%%%%%%%%%%%%%%%%%%%%%%%%%%%%%%%

\section{Exact expression for $z_k^{(3)}(a)$}
\label{app:z3}
The exact expression for $z_k^{(3)}(a)$ is given by
\be
\label{z3exact}
z_k^{(3)}(a) = \mathcal{N}/\mathcal{D}
\ee
where the numerator and denominator are respectively,
\bwt
\ber
\mathcal{N} &=& H \left(-e^{\frac{4 i k}{\sqrt{e} H}+\frac{2 i k}{H \epsilon }+\frac{5}{2}} H^4 (a H \epsilon -2 i k) \epsilon ^{3/2}+e^{\frac{2 i \left(\epsilon +\sqrt{\epsilon }\right) k}{\sqrt{e} H \epsilon }+\frac{2 i k}{H \epsilon }+\frac{5}{2}} H^4 (a H \epsilon -2 i k) \epsilon ^{3/2}-a e^{\frac{2 i \left(\sqrt{\epsilon }+2\right) k}{\sqrt{e} H \sqrt{\epsilon }}+\frac{2 (a-1) i k}{a H \epsilon }+\frac{5}{2}} H^5 \epsilon ^{5/2}\right. \nn \\
&&\left. +a e^{\frac{2 i \left(2 \epsilon +\sqrt{\epsilon }\right) k}{\sqrt{e} H \epsilon }+\frac{2 (a-1) i k}{a H \epsilon }+\frac{5}{2}} H^5 \epsilon ^{5/2}+a e^{\frac{2 i \left(2 \epsilon +\sqrt{\epsilon }\right) k}{\sqrt{e} H \epsilon }+\frac{2 (a-1) i k}{a H \epsilon }+2} H^4 i k \epsilon ^2+2 a e^{\frac{2 i \left(\sqrt{\epsilon }+2\right) k}{\sqrt{e} H \sqrt{\epsilon }}+\frac{2 (a-1) i k}{a H \epsilon }+\frac{3}{2}} H^3 k^2 \left(\sqrt{\epsilon }-1\right) \epsilon ^2\right.\nn \\
&&\left. -i a e^{\frac{2 i \left(\sqrt{\epsilon }+2\right) k}{\sqrt{e} H \sqrt{\epsilon }}+\frac{2 (a-1) i k}{a H \epsilon }+2} H^4 k \left(2 \sqrt{\epsilon }-1\right) \epsilon ^2-4 i a e^{\frac{2 i k \left(\sqrt{e} (a-1)+a \left(\epsilon +2 \sqrt{\epsilon }\right)\right)}{a \sqrt{e} H \epsilon }} k^5 \epsilon -4 a e^{\frac{2 i \left(\sqrt{\epsilon }+2\right) k}{\sqrt{e} H \sqrt{\epsilon }}+\frac{2 (a-1) i k}{a H \epsilon }+\frac{1}{2}} H k^4 \epsilon \right.\nn \\
 &&\left. -2 i a e^{\frac{2 i \left(\sqrt{\epsilon }+2\right) k}{\sqrt{e} H \sqrt{\epsilon }}+\frac{2 (a-1) i k}{a H \epsilon }+1} H^2 k^3 (\epsilon -1) \epsilon -2 i e^{\frac{2 i \left(\epsilon +\sqrt{\epsilon }\right) k}{\sqrt{e} H \epsilon }+\frac{2 i k}{H \epsilon }+1} H k^3 (a H \epsilon -2 i k) \epsilon -3 i e^{\frac{4 i k}{\sqrt{e} H}+\frac{2 i k}{H \epsilon }+2} H^3 k (a H \epsilon -2 i k) \epsilon \right. \nn\\
&&\left. -2 e^{\frac{2 i \left(\epsilon +\sqrt{\epsilon }\right) k}{\sqrt{e} H \epsilon }+\frac{2 i k}{H \epsilon }+\frac{3}{2}} H^2 k^2 \left(\sqrt{\epsilon }+1\right) (a H \epsilon -2 i k) \epsilon +e^{\frac{2 i \left(\epsilon +\sqrt{\epsilon }\right) k}{\sqrt{e} H \epsilon }+\frac{2 i k}{H \epsilon }+2} H^3 k \left(2 \sqrt{\epsilon }+1\right) (2 k+a H i \epsilon ) \epsilon \right. \nn \\
&&\left. +4 e^{\frac{4 i k}{\sqrt{e} H}+\frac{2 i k}{H \epsilon }+\frac{3}{2}} H^2 k^2 (a H \epsilon -2 i k) \sqrt{\epsilon }+2 e^{\frac{4 i k}{\sqrt{e} H}+\frac{2 i k}{H \epsilon }+1} H k^3 (2 k+a H i \epsilon )\right)
\eer
\ewt

\bwt
\begin{align}
\mathcal{D} =\,\, &4 a e^{\frac{4 i k}{\sqrt{e} H}+\frac{2 i k}{H \epsilon }+\frac{3}{2}} H^4 k^2 \epsilon ^{3/2}-e^{\frac{2 i \left(\sqrt{\epsilon }+2\right) k}{\sqrt{e} H \sqrt{\epsilon }}+\frac{2 (a-1) i k}{a H \epsilon }+\frac{5}{2}} H^5 (2 i k+a H \epsilon ) \epsilon ^{3/2}+e^{\frac{2 i \left(2 \epsilon +\sqrt{\epsilon }\right) k}{\sqrt{e} H \epsilon }+\frac{2 (a-1) i k}{a H \epsilon }+\frac{5}{2}} H^5 (2 i k+a H \epsilon ) \epsilon ^{3/2}\nn\\
&-a e^{\frac{4 i k}{\sqrt{e} H}+\frac{2 i k}{H \epsilon }+\frac{5}{2}} H^6 \epsilon ^{5/2}+a e^{\frac{2 i \left(\epsilon +\sqrt{\epsilon }\right) k}{\sqrt{e} H \epsilon }+\frac{2 i k}{H \epsilon }+\frac{5}{2}} H^6 \epsilon ^{5/2}-2 i a e^{\frac{2 i \left(\epsilon +\sqrt{\epsilon }\right) k}{\sqrt{e} H \epsilon }+\frac{2 i k}{H \epsilon }+1} H^3 k^3 \epsilon ^2-3 i a e^{\frac{4 i k}{\sqrt{e} H}+\frac{2 i k}{H \epsilon }+2} H^5 k \epsilon ^2\nn\\
&-2 a e^{\frac{2 i \left(\epsilon +\sqrt{\epsilon }\right) k}{\sqrt{e} H \epsilon }+\frac{2 i k}{H \epsilon }+\frac{3}{2}} H^4 k^2 \left(\sqrt{\epsilon }+1\right) \epsilon ^2+a e^{\frac{2 i \left(\epsilon +\sqrt{\epsilon }\right) k}{\sqrt{e} H \epsilon }+\frac{2 i k}{H \epsilon }+2} H^5 i k \left(2 \sqrt{\epsilon }+1\right) \epsilon ^2+2 a e^{\frac{4 i k}{\sqrt{e} H}+\frac{2 i k}{H \epsilon }+1} H^3 i k^3 \epsilon\nn\\
&+e^{\frac{2 i \left(2 \epsilon +\sqrt{\epsilon }\right) k}{\sqrt{e} H \epsilon }+\frac{2 (a-1) i k}{a H \epsilon }+2} H^4 i k (2 i k+a H \epsilon ) \epsilon +2 e^{\frac{2 i \left(\sqrt{\epsilon }+2\right) k}{\sqrt{e} H \sqrt{\epsilon }}+\frac{2 (a-1) i k}{a H \epsilon }+\frac{3}{2}} H^3 k^2 \left(\sqrt{\epsilon }-1\right) (2 i k+a H \epsilon ) \epsilon \nn\\
&+e^{\frac{2 i \left(\sqrt{\epsilon }+2\right) k}{\sqrt{e} H \sqrt{\epsilon }}+\frac{2 (a-1) i k}{a H \epsilon }+2} H^4 k \left(2 \sqrt{\epsilon }-1\right) (2 k-i a H \epsilon ) \epsilon -4 e^{\frac{2 i \left(\sqrt{\epsilon }+2\right) k}{\sqrt{e} H \sqrt{\epsilon }}+\frac{2 (a-1) i k}{a H \epsilon }+\frac{1}{2}} H k^4 (2 i k+a H \epsilon )\nn\\
&+4 e^{\frac{2 i k \left(\sqrt{e} (a-1)+a \left(\epsilon +2 \sqrt{\epsilon }\right)\right)}{a \sqrt{e} H \epsilon }} k^5 (2 k-i a H \epsilon )+2 e^{\frac{2 i \left(\sqrt{\epsilon }+2\right) k}{\sqrt{e} H \sqrt{\epsilon }}+\frac{2 (a-1) i k}{a H \epsilon }+1} H^2 k^3 (\epsilon -1) (2 k-i a H \epsilon)
\end{align}
\ewt

%%%%%%%%%%%%%%%%%%%%%%%%%%%%%%%%%%%%%%%%%%%%%%%%%%%%%%%%


\begin{thebibliography}{99}
\bibitem{hawk1}  S. W. Hawking, {\it Black hole explosions?,} Nature 248  30 (1974).
\bibitem{hawk2}S. W. Hawking,  {\it Particle creation by black holes,} Commun. Math. Phys. 43 (1975) 199.




\bibitem{Parker:1968mv} 
  L.~Parker,
   {\it Particle creation in expanding universes,} Phys.  Rev. Lett.  {\bf 21}, 562 (1968).

\bibitem{Parker:1968mv1}
L. Parker, Phys. Rev.  {\bf 183}, 1057 (1969).
 
 \bibitem{Parker:1968mv2} L.~Parker, {\it Quantized Fields and Particle Creation in Expanding Universes. II,} Phys. Rev. D {\bf 3}, 346 (1971) Erratum: [Phys. Rev. D {\bf 3}, 2546 (1971)].
  
  \bibitem{unruh} W. G. Unruh,  {\it Notes on black-hole evaporation,}  Phys. Rev. D {\bf 14}, 870 (1976).

%%%desitter%%%

\bibitem{Mott} E. Mottola, 
 {\it Particle Creation in de Sitter Space,} Phys. Rev. D 31 (1985).

\bibitem{GH}G. W. Gibbons and S. W. Hawking, {\it Cosmological event horizons, thermodynamics, and particle creation,} Phys. Rev. D 15, 2738 (1977).

\bibitem{ford} L. H. Ford, {\it Gravitational particle creation and inflation,} Phys. Rev. D 35, 2955 (1987). 

\bibitem{dsrev1} E.T. Akhmedov, {\it Lecture notes on interacting quantum fields in de Sitter space,} Int. J. Mod. Phys. D23 (2014) 1430001.

\bibitem{dsrev2} M. Spradlin, A. Strominger, A. Volovich, {\it Les Houches Lectures on De Sitter Space,} [arXiv:hep-th/0110007].
 
  
\bibitem{Singh:2013pxf} 
  S.~Singh, C.~Ganguly and T.~Padmanabhan,
   {\it  Quantum field theory in de Sitter and quasi-de Sitter spacetimes revisited,}
  Phys.\ Rev.\ D {\bf 87}, no. 10, 104004 (2013).

\bibitem{ads} S. Hawking, D. N. Page, {\it  Thermodynamics of black holes in anti-de Sitter space,} Comm. Math. Phys. 87, 577 (1983).
%%%desitter%%%
\bibitem{Hashiba:2018iff} 
  S.~Hashiba and J.~Yokoyama,
   {\it Gravitational reheating through conformally coupled superheavy scalar particles,}
  JCAP {\bf 1901}, no. 01, 028 (2019)


\bibitem{others2}Y. Kluger et al., {\it Pair production in a strong electric field,} Phys. Rev. Lett. 67, 2427 (1991).
\bibitem{others3}C. Kiefer et al., {\it A comparison between semiclassical gravity and semiclassical electrodynamics,} Class. Quant. Grav. 8, L185 (1991).
\bibitem{others4}C. Kiefer, {\it Functional Schrodinger equation for scalar QED,} Phys. Rev. D 45, 2044 (1992).
\bibitem{others5}A. Campos and E. Verdaguer, {\it Semiclassical equations for weakly inhomogeneous cosmologies,} Phys. Rev. D 49, 1861 (1994).
\bibitem{others6}B. L. Hu, G. Kang and A. Matacz, {\it Squeezed vacua and the quantum statistics of cosmological particle creation,} Int. J. Mod. Phys. A9, 991 (1994).
\bibitem{others7}S. Habib, C. Molina-Paris and E. Mottola, {\it Energy-momentum tensor of particles created in an expanding universe,} Phys. Rev. D 61, 024010 (2000).
\bibitem{others8}S. P. Kim, D. N. Page, {\it Schwinger pair production in electric and magnetic fields,} Phys. Rev. D 73, 065020 (2006).
\bibitem{others9}F. Cooper and G. C. Nayak , {\it Schwinger Mechanism for Fermion Pair Production in the Presence of Arbitrary Time Dependent Background Electric Field,}[arXiv:hep-th/0611125].
\bibitem{others10}S. P. Kim et al., {\it Improved approximations for fermion pair production in inhomogeneous electric fields,} Phys. Rev. D 75, 045013 (2007).


%%%effective action%%%
\bibitem{hu} E. Calzetta and B. L. Hu, {\it Nonequilibrium quantum fields: Closed-time-path effective action, Wigner function, and Boltzmann equation,} Phys. Rev. D 37, 2878 (1988).
\bibitem{buch} I.L Buchbinder, S Odintsov, L Shapiro, {\it Effective Action in Quantum Gravity,} Taylor and Francis (1992).
%%%%%effective action%%%

%%%algebraic%%%
\bibitem{ash} A. Ashtekar, Anne Magnon, and Roger Penrose, {\it Quantum fields in curved space-times,} Proceedings of the Royal Society of London. A. Mathematical and Physical Sciences, 346, 1975.

\bibitem{Kay:1988mu} 
  B.~S.~Kay and R.~M.~Wald,
   {\it Theorems on the Uniqueness and Thermal Properties of Stationary, Nonsingular, Quasifree States on Space-Times with a Bifurcate Killing Horizon,}
  Phys.\ Rept.\  {\bf 207}, 49 (1991).

\bibitem{Hollands:2004yh} 
  S.~Hollands and R.~M.~Wald,
   {\it Conservation of the stress tensor in interacting quantum field theory in curved spacetimes,}
  Rev.\ Math.\ Phys.\  {\bf 17}, 227 (2005).
  
\bibitem{hack} Thomas-Paul Hack, {\it Cosmological Applications of Algebraic Quantum Field Theory in Curved Spacetimes,} Springer, 2016.

%%%%algebraic%%%

%%%complex path%%%
\bibitem{Parikh:1999mf} 
  M.~K.~Parikh and F.~Wilczek,
    {\it Hawking radiation as tunneling,}
  Phys.\ Rev.\ Lett.\  {\bf 85}, 5042 (2000).
  
\bibitem{Parikh:1999mf1}  M.K.Parikh, {\it A secret tunnel through the horizon,} Int. J. Mod. Phys. D 13, 2351 (2004). 

\bibitem{Parikh:1999mf2} E. T. Akhmedov et al., {\it Subtleties in the quasi-classical calculation of Hawking radiation,} Int. J. Mod. Phys. D17 (2008) 2453-2458.

\bibitem{Parikh:1999mf3}E. T. Akhmedov et al., {\it Hawking temperature in the tunneling picture,}  Phys. Lett. B642 (2006) 124-128.

\bibitem{paddy} 
T. Padmanabhan, {\it Quantum theory in external electromagnetic and gravitational fields: A comparison of some conceptual issues,} Pramana, 37, 179 (1991).

\bibitem{paddy1} L. Sriramkumar and T. Padmanabhan, {\it Does a nonzero tunneling probability imply particle production in time-independent classical electromagnetic backgrounds?,} Phys. Rev. D 54, 7599 (1996).
 
\bibitem{paddy02} K.~Srinivasan and  T.~Padmanabhan, {\it Particle production and complex path analysis,} Phys. Rev. D 60, 024007 (1999).

\bibitem{paddy3} S. Shankarnarayanan, K.Srinivasan and T.Padmanabhan, {\it Method of complex paths and general covariance of Hawking radiation,} Mod. Phys. Lett. A 16, 571 (2001).
\bibitem{paddy4} E. Vagenas, {\it Complex paths and covariance of Hawking radiation in 2D stringy black holes,} Nuovo Cim. B 117, 899 (2002).


%%%complexpath%%%
%%%%usds%%%
\bibitem{usds}
D. Boyanovsky, et al.   {\it Particle decay during inflation: Self-decay of inflaton quantum fluctuations during slow roll,} Phys.Rev. D71 (2005) 023509 2815-2823.

\bibitem{usds1}A.M. Polyakov,  {\it De Sitter space and eternity,} Nucl. Phys. B 797 (2008) 199-217. 

\bibitem{usds2} A.M. Polyakov,  {\it Decay of Vacuum Energy,} Nucl. Phys. B834 (2010) 316-329. 

\bibitem{usds3} A. Higuchi,  {\it Decay of the free-theory vacuum of scalar field theory in de Sitter spacetime in the interaction picture,}  Class. Quant. Grav. 26 (2009) 072001. 

\bibitem{usds4}E.T Akhmedov et al.,  {\it De Sitter space and perpetuum mobile,}  Phys. Atom. Nucl. 75 (2012) 525-529.

\bibitem{usds5}E.T. Akhmedov, et al., {\it Interacting Field Theories in de Sitter Space are Non-Unitary,}  Phys. Rev. D78 (2008) 104005. 

\bibitem{usds6} E.T. Akhmedov, et al. {\it A Simple way to take into account back reaction on pair creation,} Phys. Lett. B687 (2010) 267-270. 

\bibitem{usds7}E.T. Akhmedov, {\it Real or Imaginary? (On pair creation in de Sitter space),} Mod.Phys.Lett. A25 (2010).

\bibitem{usds8}P.~R.~Anderson, E.~Mottola and D.~H.~Sanders, {\it Decay of the de Sitter Vacuum,} Phys.\ Rev.\ D {\bf 97}, no. 6, 065016 (2018). 

\bibitem{usds9}H.~Matsui, {\it Instability of De Sitter Spacetime induced by Quantum Conformal Anomaly,} JCAP {\bf 1901}, no. 01, 003 (2019).

\bibitem{usds10}A.~Rajaraman, {\it  de Sitter Space is Unstable in Quantum Gravity,}
  Phys.\ Rev.\ D {\bf 94}, no. 12, 125025 (2016).
  
\bibitem{usds11}  C.~M.~Ho and S.~D.~H.~Hsu,
    {\it  Instability of Quantum de Sitter Spacetime,} JHEP {\bf 1504}, 086 (2015).
   
\bibitem{usds12}   P.~R.~Anderson and E.~Mottola, {\it Quantum vacuum instability of ``eternal'' de Sitter space,} Phys.\ Rev.\ D {\bf 89}, 104039 (2014). 

\bibitem{usds13}V.~Emelyanov and F.~R.~Klinkhamer,
   {\it  De Sitter-spacetime instability from a nonstandard vector field,}
  Phys.\ Rev.\ D {\bf 86}, 027302 (2012).


\bibitem{sds}
E. Alvarez, et al. {\it  Eternity and the cosmological constant,}  JHEP 0910 (2009) 045.

\bibitem{sds1}D. Marolf, et al. {\it  The IR stability of de Sitter: Loop corrections to scalar propagators,}  Phys. Rev. D82 (2010) 105032. 

\bibitem{sds2} D. Marolf, et al. {\it The IR stability of de Sitter QFT: Physical initial conditions,}  Gen.Rel.Grav. 43 (2011) 3497-3530.

\bibitem{sds3} A. Higuchi, et al. {\it On the Equivalence between Euclidean and In-In Formalisms in de Sitter QFT,}  Phys. Rev. D83 (2011) 084029.

\bibitem{sds4} D. Marolf, et al. {\it The IR stability of de Sitter QFT: results at all orders,}  Phys. Rev. D84 (2011) 044040.  

\bibitem{sds5}G.~Moreau and J.~Serreau,
  {\it Stability of de Sitter spacetime against infrared quantum scalar field fluctuations,}
  Phys.\ Rev.\ Lett.\  {\bf 122}, no. 1, 011302 (2019).

%reviews

\bibitem{DeWitt:1975ys} 
  B.~S.~DeWitt,
   {\it  Quantum Field Theory in Curved Space-Time,}
  Phys.\ Rept.\  {\bf 19}, 295 (1975).
  
  \bibitem{Brout:1995rd} 
  R.~Brout, S.~Massar, R.~Parentani and P.~Spindel,
   {\it  A Primer for black hole quantum physics,}
  Phys.\ Rept.\  {\bf 260}, 329 (1995)
  
  \bibitem{Brandenberger:1984cz} 
  R.~H.~Brandenberger,
    {\it  Quantum Field Theory Methods and Inflationary Universe Models,}
  Rev.\ Mod.\ Phys.\  {\bf 57}, 1 (1985).


\bibitem{rev-dun} 
G. V. Dunne, in {\it  From Fields to Strings: Circumnavigating Theoretical Physics,} Shifman, M. (ed.) et al. (2004) [arXiv:hep-th/0406216].

\bibitem{rev-pad1}T. Padmanabhan, {\it Gravity and the thermodynamics of horizons,} Phys. Rept. 406, 49 (2005).

\bibitem{rev-pag1}D. N. Page, {\it Hawking radiation and black hole thermodynamics,} New J. Phys. 7, 203 (2005) [arXiv:hep-th/0409024].


\bibitem{rev-muk} V. F. Mukhanov, H. A. Feldman and R. H. Brandenberger, {\it Theory of cosmological perturbations,} Phys. Rept. 215, 203 (1992).

\bibitem{rev-mar}
J. Martin, {\it Inflationary cosmological perturbations of quantum-mechanical origin,} Lect. Notes Phys. 669, 199 (2005).


%%%%reviews%%%


%%%books%%%
\bibitem{birrel} N. D. Birrel and P. C. W. Davies, {\it Quantum fields in Curved Space,} Cambridge Monographs on Mathematical Physics, Cambridge University Press, 1982.

\bibitem{Fulling:1989nb} 
  S.~A.~Fulling,
  {\it Aspects of Quantum Field Theory in Curved Space-time,}
  London Math.\ Soc.\ Student Texts {\bf 17}, 1 (1989).
  
  
  \bibitem{book-pad} T. Padmanabhan, {\it Theoretical Astrophysics, Volume III: Galaxies and Cosmology,} (Cambridge University Press, Cambridge, England, 1999), Sec. 5.7.
  

  
 \bibitem{Mukhanov:2007zz} 
  V.~Mukhanov and S.~Winitzki,
    {\it Introduction to quantum effects in gravity,} Cambridge University Press, 2007.
   
\bibitem{parker}  L. Parker and D. Toms,  {\it Quantum Field Theory in Curved Spacetime: Quantized Fields and Gravity,} Cambridge Monographs on Mathematical Physics, 2009.  

\bibitem{paddy2} T.~Padmanabhan, {\it Gravitation: Foundations and Frontiers}, Cambridge University Press, 2010.   


%%%books%%%

%\bibitem{modak} S. K. Modak et al, In Progress.

\bibitem{modak1} 
  S.~K.~Modak,
    {\it New geometric and field theoretic aspects of a radiation dominated universe,}
  Phys.\ Rev.\ D {\bf 97}, no. 10, 105016 (2018).
 
 \bibitem{modak2} 
  S.~K.~Modak,
    {\it New geometric and field theoretic aspects of a radiation dominated universe II. Fundamental Cosmological Observers (FCOs),}
  [arXiv:1806.00972 [gr-qc]].


\bibitem{fulling} S.A. Fulling and P. C. W. Davies, {\it Radiation from a moving mirror in two dimensional space-time: conformal anomaly,} Proc. R. Soc. London, {\bf A348} 393 (1976).

\bibitem{davies}  P. C. W. Davies and S.A. Fulling, {\it Radiation from moving mirrors and from black holes,} Proc. R. Soc. London, {\bf A356} 237 (1977).


 \bibitem{schwinger}
J. Schwinger, {\it On gauge invariance and vacuum polarization,} Phys. Rev. 82, 664 (1951).


\bibitem{othr}S. W. Hawking, {\it The development of irregularities in a single bubble inflationary universe,} Phys. Lett. B 115, 295 (1982).

\bibitem{othr1}
A. A. Starobinsky, {\it Dynamics of phase transition in the new inflationary universe scenario and generation of perturbations,} Phys. Lett. B 117, 175 (1982).

\bibitem{othr2} A. H. Guth and S.-Y. Pi, {\it Fluctuations in the new inflationary universe,} Phys. Rev. Lett. 49, 1110 (1982).

\bibitem{othr3} A. D. Dolgov, A. D. Linde, {\it Baryon asymmetry in the inflationary universe,} Phys. Lett. B 116, 329 (1982).

\bibitem{othr4} J. M. Bardeen, P. J. Steinhardt and M. S. Turner, {\it Spontaneous creation of almost scale-free density perturbations in an inflationary universe,} Phys. Rev. D 28, 679 (1983).

\bibitem{othr5} L. F. Abbott and M. B. Wise, {\it Constraints on generalized inflationary cosmologies,} Nucl. Phys. B 244, 541 (1984).

\bibitem{othr6} T. Padmanabhan, {\it Acceptable density perturbations from inflation due to quantum gravitational damping,} Phys. Rev. Lett. 60 2229 (1988). 

\bibitem{othr7} T. Padmanabhan, T. R. Seshadri and T. P. Singh, {\it Making inflation work: Damping of density perturbations due to Planck energy cutoff,} Phys. Rev. D 39, 2100 (1989).

\bibitem{lyth2008}
D.~H.~Lyth and D.~Seery, {\it Classicality of the primordial perturbations,} Phys. Lett. B, 662, 309 (2008). 


\bibitem{gaurang2007}
G.~Mahajan and T.~Padmanabhan,  {\it Particle creation, classicality and related issues in quantum field theory: I. Formalism and toy models,} Gen.~Rel.~Grav., 40, 661 (2008).

\bibitem{gaurang20071}
G.~Mahajan and T.~Padmanabhan, {\it Particle creation, classicality and related issues in quantum field theory: II. Examples from field theory,} Gen.~Rel.~Grav., 40, 709 (2008).


\bibitem{gaurang2008}
G.~Mahajan, {\it Particle creation in a time-dependent electric field revisited,} Annals. Phys. 324, 361 (2009). 


\bibitem{sriram}
L. Sriramkumar and T. Padmanabhan, {\it Initial state of matter fields and trans-Planckian physics: Can CMB observations disentangle the two?,} Phys. Rev. D 71, 103512 (2005).  



\bibitem{Singh:2013bsa} 
  S.~Singh, S.~K.~Modak and T.~Padmanabhan,
  {\it Evolution of quantum field, particle content and classicality in the three stage universe,}
  Phys.\ Rev.\ D {\bf 88}, no. 12, 125020 (2013).


\bibitem{cosmin}
T. Padmanabhan, {\it Emergent perspective of gravity and dark energy,} Res. in Astro. Astrophys.,12, 891 (2012).

\bibitem{cosmin1}Hamsa Padmanabhan, T. Padmanabhan, {\it CosMIn: The solution to the cosmological constant problem,} Int. Jour. Mod. Phys, D 22, 1342001 (2013).  



\bibitem{bunchdavies}
T.~S.~Bunch and P.~C.~W~Davies, {\it Non-conformal renormalised stress tensors in Robertson-Walker space-times,} J. Phys. A, 11, 1315 (1978).
 
  



\end{thebibliography}
\end{document}